\documentclass[reprint,
amsmath, 
amssymb,
aps, 
superscriptaddress,
prb,
nobibnotes
]
{revtex4-1}

\usepackage{array}
\usepackage{microtype}

\usepackage{graphicx}
\usepackage{amssymb}
\usepackage{bm,dsfont}
\usepackage{multirow}
\usepackage{color}
\usepackage{amsmath}
\usepackage{threeparttable}
\usepackage[nolist,nohyperlinks]{acronym}
\usepackage{hyperref}
\usepackage[nameinlink,capitalize]{cleveref} % For Equations
\usepackage{placeins}
\usepackage[normalem]{ulem}

\usepackage{mathrsfs}
\usepackage{enumitem}

\newcommand{\rev}[1]{{\color{black} #1}}
\newcommand{\rdel}[1]{}

\newcommand{\ue}{\mathrm{e}}
\newcommand{\ud}{\mathrm{d}}
\newcommand{\ui}{\mathrm{i}}

\newcommand{\Rg}{\mbox{$R_\text{g}$}}
\newcommand{\Rgcube}{\mbox{$R^3_\text{g}$}}
\newcommand{\Ree}{\mbox{$R_\text{ee}$}}
\newcommand{\ie}{\textit{i.e.}}

\newcommand{\kBT}{k_{_\textrm{B}}\!T}

\newcommand{\lhat}{\hat{\Lambda}}
\newcommand{\llhat}{\hat{\bm{\Lambda}}}
\newcommand{\bg}{\bm{g}}
\newcommand{\bG}{\bm{G}}

\newcommand{\muhat}{\hat{\mu}}
\newcommand{\jhat}{\hat{j}}
\newcommand{\jjhat}{\hat{\vec{j}}}
\newcommand{\rr}{\vec{r}}
\newcommand{\qq}{\vec{q}}
\newcommand{\uu}{\vec{u}}
\newcommand{\RR}{\vec{R}}

\newcommand{\vzeta}{\vec{\zeta}}
\newcommand{\vtheta}{\vec{\theta}}

\newcommand{\vH}{\vec{H}}

\newcommand{\Sgrid}{S_\textrm{grid}}
\newcommand{\sgrid}{\sigma_\textrm{grid}}
\newcommand{\ssgrid}{\sigma_\textrm{grid}^2}
\newcommand{\tO}{t_0}

 \newcommand{\GG}{\bm{G}}
 \newcommand{\II}{\bm{I}}
 \newcommand{\ee}{\bm{e}}
 \newcommand{\cc}{\bm{c}}
 \newcommand{\CC}{\bm{C}}

 \newcommand{\subsum}{{\sum_{n \in S_\alpha \atop m \in S_\beta}}}

 \newcommand{\Dc}{D_\textrm{c}}

 \newcommand{\xemat}[4]{
       \left(
           \begin{array}{c|c}
               #1 & #2
           \\  \hline #3 & #4
           \end{array}
       \right)
       }
 \newcommand{\mat}[1]{\underline{\underline{\bar{#1}}}}
 \newcommand{\vvec}[1]{\underline{\bar{#1}}}

\begin{document}

\title{From Single-Chain Dynamics to Structure Formation:
Dynamic Self-Consistent Field Theory and Molecular Dynamics 
of (Co)polymer \rdel{m}\rev{M}elts across Entanglement Regimes}

\author{Alireza F. Behbahani}
 \email{aforooza@uni-mainz.de}
 \affiliation{Institut f\"{u}r Physik, 
   Johannes Gutenberg-Universit\"{a}t Mainz, 
   Staudingerweg 7, D-55099 Mainz, Germany}

\author{Jafar Cheraghalizadeh}
 \email{jcheragh@uni-mainz.de}
 \affiliation{Institut f\"{u}r Physik, 
   Johannes Gutenberg-Universit\"{a}t Mainz, 
   Staudingerweg 7, D-55099 Mainz, Germany}

 \author{Friederike Schmid}
 \email{friederike.schmid@uni-mainz.de}
 \affiliation{Institut f\"{u}r Physik, 
   Johannes Gutenberg-Universit\"{a}t Mainz, 
   Staudingerweg 7, D-55099 Mainz, Germany}

%\begin{document}   

\begin{abstract}
  \begin{center}
%\textbf{Abstract}
\end{center}
Dynamic self-consistent field theory (DSCFT) provides an efficient
continuum framework for studying structure formation in inhomogeneous
polymer systems, but its predictive accuracy depends on the choice of
the nonlocal mobilities. Here, we construct mobility functions for
moderately and strongly entangled homopolymer and diblock copolymer
systems from the relaxation dynamics of single-chain structure factors,
based on molecular dynamics (MD) simulations of the Kremer-Grest model
and analytical reptation theory. Single-chain mobilities are combined
such that the resulting DSCFT accounts for the dependence of fluxes on
local chain densities.  The theory is then applied to the spinodal
decomposition of symmetric homopolymer blends and diblock copolymer
melts following a quench into the (micro)phase-separation regime.
Predictions of DSCFT are systematically compared with MD simulations.
Mobility functions derived from single-chain dynamics are found to
reproduce the kinetics of structure formation more accurately than
conventional Debye-type mobilities. We additionally investigate the
influence of adding stochastic currents (noise) that are correlated
according to the fluctuation–dissipation relation. At low noise levels,
they enable the generation of equilibrium initial states and facilitate
defect annealing. At high noise levels, however, nonlinear effects lead
to discrepancies between DSCFT and MD simulations.
     
\end{abstract}
 
 \maketitle
 
\section{Introduction}  

\label{sec:intro}

Polymer blends and block copolymers, composed of two or more
chemically distinct monomeric units, have been extensively studied
because of their broad range of applications.  The entropic driving
force for mixing of long polymer chains is inherently weak; according
to the Flory-Huggins theory, the entropy of mixing per unit volume
scales inversely with the degree of polymerization, \ie, the chain
length $N$. Consequently, even a slight incompatibility between
different monomer species can lead to immiscibility of
polymers~\cite{flory1953principles}. In blends, immiscible polymers
phase separate into domains enriched in one component or the other.
Block copolymers exhibit more complex behavior; the covalent linkage
between chemically distinct blocks prevents macroscopic phase
separation, leading instead to the formation of ordered
microstructures such as lamellae, hexagonally packed cylinders,
body-centered cubic spheres, or the bicontinuous gyroid phase. Such
phenomena can be described theoretically using the self-consistent
field theory (SCFT), a powerful framework for predicting the structure
and thermodynamics of heterogeneous polymer
systems~\cite{matsen1994stable,schmid1998,
fredrickson2006equilibrium}.                                              

When a block copolymer melt or a polymer blend is brought into a
thermodynamically unstable state, the system spontaneously evolves to
reduce its free energy. However, such systems may remain out of
thermodynamic equilibrium for extended periods -- first, because
polymer dynamics are intrinsically slow and span a broad range of time
scales, and second, because the system can become trapped in local
minima of the free energy landscape~\cite{muller2013computational}.
This latter effect may give rise to long-lived defects and metastable
intermediate morphologies during copolymer
self-assembly~\cite{takahashi2012defectivity,raybin2017real}.
Understanding the kinetics of structure formation is therefore
crucial for controlling and designing processing routes used to
prepare polymer materials with desired morphologies. 

An important approach to studying the dynamics of structure formation is
through particle-based simulations of coarse-grained polymer models.
However, such simulations are often computationally demanding, as they
typically require large system sizes, many particles, and long
simulation times.  This challenge is particularly severe in 
inhomogeneous entangled polymer melts, where the relaxation time of
entangled chains increases strongly with chain length ($\sim
N^{3.4}$)~\cite{mcleish2002tube}.  Consequently, simulations of very
long, highly entangled chains remain computationally expensive even when
highly coarse-grained soft polymer models are employed, with
entanglement constraints represented by
slip-springs~\cite{masubuchi2006primitive,ramirez2018detailed,behbahani2021dynamics}.
Another limitation of particle-based simulations is that they do not
provide direct access to the free energies of system configurations
during structure formation. Such information is important because
knowledge of free energies can provide valuable insight into kinetic
pathways, for example, by identifying metastable configurations
associated with local minima of the free energy
landscape~\cite{takahashi2012defectivity}.

An alternative, computationally efficient approach to model the
dynamics of structure formation at the continuum level is to use the
frame work of dynamic self-consistent field theory (DSCFT), which
incorporates the SCFT free energy functional in a dynamic density
functional equation~\cite{fraaije1997dynamic,reister2001spinodal,
muller2005incorporating}.  The general equation describing the
diffusive time evolution of the local density fields $\rho_\alpha$ of
monomer species $\alpha$ in a (co)polymer melt
is~\cite{qi2017dynamic}:
\begin{equation}
\frac{\partial \rho_{\alpha}(\rr,t)}{\partial t}  
= \nabla \cdot 
\sum_{\beta} \int \ud^3 r'
\Lambda_{\alpha \beta}(\rr,\rr') 
   \nabla \mu_{\beta}(\rr',t),
\label{Eq:DDFT-rho}
\end{equation}
where the sum runs over monomer species \rdel{$\alpha,$} $\beta$.  Here
$\mu_{\beta}(\rr',t) = \delta {\cal F}/\delta \rho_{\beta}(\rr',t)$ is
the excess chemical potential derived from the SCFT free energy
functional ${\cal F}[\rho_\alpha]$, and $\Lambda_{\alpha
\beta}(\rr,\rr')$ is a nonlocal mobility matrix connecting thermodynamic
forces at $\rr'$ to fluxes at $\rr$. The non-local structure of
$\Lambda$ reflects the \rdel{of} chain connectivity, which enables propagation
of thermodynamic forces along the chain ~\cite{de1981coherent,
binder2001spinodal, muller2013computational}, and also encodes
information on the chain dynamics, such as Rouse or reptation
behavior~\cite{de1981coherent}. Although the mobility matrix is
nonlocal, it is taken to be frequency independent, implying that memory
effects are neglected.  Dynamic density functional approaches that
incorporate memory effects have also been proposed
recently~\cite{wang2019collective, rottler2020kinetic}. 

Following up on early suggestions by de Gennes, Pincus, and Binder
~\cite{de1980dynamics, de1981coherent, pincus1981dynamics, binder1983collective}, our
group recently proposed computing the mobility matrix from the mean
wave-vector dependent relaxation time of the single-chain dynamic
structure factor, $g(q,t)$, in homogeneous reference
system~\cite{mantha2020bottom}. For unentangled Rouse chains, we
determined mobility functions for homopolymer blends and multiblock
copolymers based on both particle-based simulations and analytical
expressions for $g(q,t)$~\cite{schmid2020dynamic}.  One-dimensional
DSCFT calculations for copolymer melts employing these mobility
functions showed excellent agreement with particle-based
simulations~\cite{mantha2020bottom}.  More recently, we calculated
mobility functions for homopolymer chains from simulations of $g(q,t)$ 
across chain lengths ranging from unentangled to moderately entangled
regimes~\cite{behbahani2024relaxation}.  

The purpose of the present work is three-fold. First, we generalize the
previous homopolymer calculation to diblock copolymers and to the highly
entangled regime: We use molecular dynamics simulations to calculate and
analyze the mobility matrix for melts of unentangled and moderately
entangled diblock copolymer chains. Additionally, we analytically
determine the mobility matrix for very long, highly entangled
homopolymer and block copolymer chains using the reptation model.
Second, we apply the resulting DSCFT equations to calculate the dynamics
of spinodal decomposition in homopolymer blends and copolymer melts
after a quench, and compare the DSCFT predictions to molecular dynamics
simulations of Kremer-Grest type chains. Among other, this also entails
generalizing our DSCFT Ansatz to nonuniform polymer mixtures, taking
into account that local fluxes should depend on local densities.
Third, we analyze the effect of stochastic currents
\cite{fraaije1997dynamic,muller2005incorporating}, which are sometimes
added in dynamic density functional theories\cite{illien2025dean} to
mimic thermal fluctuations and enable free energy barrier crossing.  We
examine their role in preparing initial configurations and their impact
on the subsequent dynamics of the structure formation.

The manuscript is organized as follows: In the next section, we
provide further details about the DSCFT and present the extension of
our approach to blends. In the following sections, we first calculate
and discuss the mobility functions, then compare the results of DSCFT
calculations for spinodal decomposition in homopolymer blends and
copolymer melts with those from molecular dynamics simulations, and finally
study the effect of stochastic fluctuations. We close with a summary
and conclusion.

\section{Dynamical self-consistent field theory}
\label{sec:ddft-intro}

The starting point of the DSCFT is a density functional ${\cal
F}[\rho_\alpha]$ describing the free energy of a system with local
densities $\rho_\alpha(\rr)$ of monomer species $\alpha$ in a fixed
volume $V$.  In the present work, we use the standard self-consistent
field expression for dense melts of Gaussian chains $\gamma$ with
Flory-Huggins type density-dependent interactions, which we write in
reduced form as ${\cal F}[\rho_\alpha] = (\kBT \rho_0/N) \:
F[\phi_\alpha]$,  where $1/\kBT$ is the Boltzmann factor, $\rho_0$ is
the mean monomer density, and $N$ is a reference chain length. The reduced free energy function depends on the
normalized dimensionless density fields $\phi_\alpha(\rr) =
\rho_\alpha(\rr)/\rho_0$ and has the form~\cite{schmid1998,fredrickson2006equilibrium}
\begin{equation}
\label{eq:energy-scf}
F[\phi_\alpha]= U[\phi_\alpha]
- \sum_\alpha \! \int \!\! \ud^3 r
  \: \phi_\alpha(\rr) w_\alpha(\rr)
- \sum_\gamma \frac{n^{(\gamma)}}{\rho_0/N} \ln({\cal Q}_\gamma).
\end{equation}
The functional $U[\phi_\alpha]$ describes the non-bonded interactions
between monomers, $n^{(\gamma)}$ is the number of chains of type
$\gamma$, and ${\cal Q}^{(\gamma)}$ the single chain partition
function for non-interacting chain species $\gamma$ subject to the
potential $w_\alpha(\rr)$, which is determined self-consistently such
that the density of the non-interacting chain system reproduces
$\phi_\alpha(\rr)$.  \cref{eq:energy-scf} can be derived as the
mean-field limit of a fluctuating field theory for interacting polymer
systems, and the dimensionless Ginzburg parameter $G = \rho_0
\Rgcube/N$ sets the inverse scale of the fluctuations that have been
neglected ($\Rg$ is the gyration radius of an ideal chain with length
$N$).  Higher values of $G$ correspond to smaller fluctuations.

In our applications, we consider (co)polymers consisting of
monomer species A and B with the interaction potential
\begin{equation}
\label{eq:flory_huggins}
U[\phi_\alpha]
= \int \! \ud^3 r \left\{ \chi N \phi_A \phi_B
+ \frac{\kappa N}{2} \big(\sum_{\alpha} \phi_\alpha - 1\big)^2 
\right\}.
\end{equation}
The first term describes the monomer incompatibility, which is quantified 
by the Flory Huggins parameter $\chi$, and the second term ensures
$\sum_\alpha \phi_\alpha \approx 1$ everywhere 
($\kappa$ is an inverse compressibility).  

In terms of the rescaled quantities $F$ and $\phi$, the DSCFT
equation, \cref{Eq:DDFT-rho}, can be rewritten as 
\begin{equation}
\frac{\partial \phi_{\alpha}(\rr,t)}{\partial t} 
  = \nabla \cdot 
%\left( 
     \sum_{\beta} \int \ud^3 r' \: \lhat_{\alpha \beta}(\rr,\rr') 
    \nabla \muhat_{\beta}(\rr',t)
%    +\hat{\mathbf{j}}_{\alpha}(\mathbf{r},t)\right) \,
\label{Eq:DDFT-phi-deterministic}
\end{equation}
with $\muhat_{\beta} = \delta F/\delta \phi_{\beta} = \frac{N}{\kBT}
\: \mu_{\beta}$, and $\lhat_{\alpha \beta} = 
\frac{\kBT}{\rho_0 N} \: \Lambda_{\alpha \beta}$.  
The resulting dynamics is deterministic and purely downhill, \ie, the
value of the free energy functional decreases monotonically with time.
Therefore, the theory cannot capture thermally activated processes
such as nucleation. Moreover, it does not account for thermal density
fluctuations in homogeneous states and therefore cannot be used to
calculate collective structure factors.  To overcome these limitations
and introduce fluctuations at least at the level of the random phase
approximation, Fraaije and coworkers\cite{fraaije1997dynamic} proposed
to introduce fluctuating currents in \cref{Eq:DDFT-phi-deterministic}.
One obtains the stochastic dynamic density functional
equation\cite{fraaije1997dynamic,reister2001spinodal,
muller2005incorporating}
\begin{equation}
\frac{\partial \phi_{\alpha}(\rr,t)}{\partial t} 
  = \nabla \cdot 
\left\{ 
     \sum_{\beta} \int \!\! \ud^3 r' \: \lhat_{\alpha \beta}(\rr,\rr') 
    \nabla \muhat_{\beta}(\rr',t)
    +\jhat_{\alpha}(\rr,t)\right\} \,
\label{Eq:DDFT-phi}
\end{equation}
(\^Ito interpretation \cite{illien2025dean}), which includes
a random current $\jhat_{\alpha}(\rr,t)$. The three spatial
components of $\jhat_{\alpha}(\rr,t)$ are Gaussian random variables 
with zero mean and correlations obeying a fluctuation–dissipation 
relation:
\begin{equation}
    \langle \jjhat_{\alpha}(\rr, t) \jjhat_{\beta}(\rr', t') \rangle 
      = 2 \: \sigma^2 \: \Rgcube \: \mathds{1} \: \delta(t - t') \: 
        \lhat_{\alpha \beta}(\rr, \rr')
\label{Eq:noise-fluc-dis}
\end{equation}
($\jjhat \jjhat$ denotes a tensor product).
The dimensionless parameter $\sigma^2$ sets the amplitude of the
noise. The most natural choice is $\sigma^2 = 1/G$; however, we shall
see below that this may not be appropriate if $G$ is small, since
the stochastic DSCFT approach becomes problematic for large
$\sigma^2$.

If the mobility is translationally invariant, 
$\lhat_{\alpha \beta}(\rr, \rr') =
  \lhat_{\alpha \beta}(\rr - \rr')$, then
\cref{Eq:DDFT-phi} can be simplified by Fourier transform 
(defined as $f(\qq) =  \int_V \ud^3 r \, 
\ue^{\ui \, \qq \cdot \rr} f(\rr)$), giving
 \begin{equation}
  \frac{\partial \phi_\alpha(\qq,t)}{\partial t} 
  = -q^2 \sum_\beta \lhat_{\alpha \beta}(\qq)\:\muhat_\beta(\qq)
    + i \:\qq \cdot \jjhat_\alpha (\qq, t)
\label{Eq:DDFT-q}
\end{equation}
with the fluctuation-dissipation relation
\begin{equation}
\langle\jjhat_{\alpha}(\qq, t) \jjhat_{\beta}(\qq', t') \rangle  
  = 2 \sigma^2 \Rgcube \: \mathds{1} \: \delta(t - t') 
    \delta_{\qq,-\qq'} \lhat_{\alpha \beta}(\qq) \: V.
\label{Eq:noise-q}
\end{equation}
To produce such a correlated random noise in practice, we first
determine a matrix $A_{\alpha\gamma}$ that obeys
\begin{equation}
\label{Eq:A_alph_gam}
\sum_{\gamma} A_{\alpha\gamma}(\qq) \: A_{ \beta \gamma}(\qq)
   = q^2 \lhat_{\alpha \beta}(\qq)  
\end{equation}
by Cholesky decomposition, then generate uncorrelated
three-dimensional Gaussian white noise vector fields
$\vzeta_\alpha(\rr,t)$ in real space ($\langle \vzeta \rangle \equiv
0$, $\langle \vzeta(\rr,t) \vzeta(\rr',t') \rangle = \mathds{1}
\delta(\rr - \rr') \delta(t - t')$), Fourier transform them to obtain
$\vzeta(\qq,t)$, and finally calculate $\jjhat_{\alpha}(\qq) = \sqrt{2
\Rgcube} \sigma \sum_ {\gamma} A_{\alpha\gamma} \vzeta_{\gamma}(\qq)$.

In the case of copolymer melts, assuming translationally invariant
mobility functions is well-justified. Polymer blends, however, may
exhibit spatially nonuniform chain densities for different polymer
types $\gamma$; consequently, the flux at a given point should be
proportional to the local chain density. To take this into account, we
propose to model the dynamic evolution $\partial_t
\rho_\alpha^{(\gamma)}$ of monomer species $\alpha$ in chains of type
$\gamma$ separately for each chain type, using a modified mobility
function
\begin{equation}
\Lambda_{\alpha\beta}^{(\gamma)}(\rr,\rr';t) 
    = \frac{1}{N_{\gamma}} \overline{\rho^{(\gamma)}(\rr,\rr';t)}
      \: \Lambda_{\alpha\beta}^{(s,\gamma)}(\rr,\rr').
\label{Eq:flux-local-rho}
\end{equation}
Here $N_\gamma$ is the length of $\gamma$-chains,
$\Lambda^{(s,\gamma)}_{\alpha \beta}$ represents the mobility {\em per 
chain} in a pure one-component reference melt of $\gamma$-chains
($\Lambda^{(s,\gamma)}_{\alpha \beta}
  = \Lambda^{(\gamma)\text{melt}}_{\alpha \beta}
 \frac{N_\gamma}{\rho_0}$), and 
 $\overline{\rho^{(\gamma)}(\rr, \rr';t)}$ is a locally 
averaged density of monomers that are part of $\gamma$-chains,
which must be symmetric with respect to interchange of $\rr$ 
and $\rr'$. We propose to approximate it by the geometric mean,
\begin{equation}
\overline{\rho^{(\gamma)}(\rr, \rr';t)} \approx
\sqrt{\rho^{(\gamma)}(\rr,t)\rho^{(\gamma)}(\rr',t)}.
\end{equation}
\rev{This multiplicative form is mostly chosen for numerical 
convenience, to keep the computational costs of evaluating
the stochastic currents at a minimum. }
Defining the scaled single-chain mobility 
%$\lhat^{(s,\gamma)}_{\alpha \beta}$ via 
  $\lhat^{(s,\gamma)}_{\alpha \beta} 
= \frac{\kBT}{N_\gamma^2} \Lambda^{(s,\gamma)}_{\alpha \beta}
  = \frac{\kBT}{\rho_0 N_\gamma} 
      \Lambda^{(\gamma)\text{melt}}_{\alpha \beta}$,
we finally obtain modified rescaled DSCFT equations for
each density field $\phi_\alpha^{(\gamma)}(\rr,t)$.
They have the same form than \cref{Eq:DDFT-phi}, but
the mobility functions now depend on the density:
\begin{equation}
\lhat^{(\gamma)}_{\alpha \beta}(\rr,\rr';t)
 = \frac{N_\gamma}{N} 
   \sqrt{\phi^{(\gamma)}(\rr,t)}
    \: \lhat^{(s,\gamma)}_{\alpha,\beta}(\rr - \rr') \:
   \sqrt{\phi^{(\gamma)}(\rr',t)},
\label{Eq:mobility-flux-local-phi}
\end{equation}
with $\phi^{(\gamma)}(\rr,t) = \sum_\alpha
\phi_\alpha^{(\gamma)}(\rr,t)$. 

These equations are again conveniently solved in Fourier
space, replacing \cref{Eq:DDFT-q} by
\begin{equation}
\partial_t \phi_\alpha^{(\gamma)}(\qq,t)  = 
i \qq \cdot \left\{
\frac{N_\gamma}{N} \vH_\alpha^{(\gamma)}(\qq,t)
+ \jjhat_\alpha^{(\gamma)}(\qq,t) \right\},
\end{equation}
where $\vH_\alpha^{(\gamma)}(\qq,t)$ is calculated from
successive Fourier transforms $\mathscr{F}$ and inverse
Fourier transforms $\mathscr{F}^{-1}$ via
\begin{displaymath}
 \begin{split}
 \vH_\alpha^{(\gamma)}(\qq,t) = &
 \mathscr{F} \big[\sqrt{\phi^{(\gamma)}(\rr,t)} \:
 \mathscr{F}^{-1} \big[ \sum_\beta \lhat_{\alpha \beta}^{(s,\gamma)} (\qq) 
 \times
 \\ & \hspace*{4mm}
 \mathscr{F}\big[ \sqrt{\phi^{(\gamma)}(\rr,t)} \: 
 \mathscr{F}^{-1} \big[i \qq  \nabla \muhat_\alpha(\qq,t)) \big]
 \: \big] \: \big] \: \big]. 
 \end{split}
\end{displaymath}
To generate the correlated stochastic current
$\jjhat_\alpha^{(\gamma)}$, we again start from uncorrelated
Gaussian white noise and calculate $\vtheta_\alpha^{(\gamma)}(\qq,t) =
\sqrt{2 \Rgcube} \sigma \sum_\beta A_{\alpha \beta}(\qq,t)
\vzeta_\beta^{(\gamma)}(\qq)$ as described previously, but then 
perform one more step in real space to get
$\jjhat_\alpha^{(\gamma)}(\rr,t) =
\sqrt{\frac{N_\gamma}{N}\phi^{(\gamma)}(\rr)} \:
\vtheta_\alpha^{(\gamma)}(\rr,t)$.

In the remaining paper, we will either consider AB copolymer melts 
or blends of homopolymers A and B with equal length $N$. In these
cases, the DSCFT can be cast in the form of \cref{Eq:DDFT-phi}
with $\lhat_{\alpha \beta}(\rr,\rr') = 
\lhat_{\alpha \beta}^{(s,\text{cop})}(\rr - \rr')$ 
for copolymers, and
\begin{equation}
\lhat_{\alpha \beta} (\rr,\rr';t)= \delta_{\alpha \beta} \:
\lhat^{(s,\text{homo})}(\rr - \rr') \:
\sqrt{\phi_\alpha(\rr,t) \: \phi_\beta(\rr',t)} 
\end{equation}
for homopolymer blends. In the next section, we will focus on
the single-chain mobility functions that enter these expressions.

\section{Single-Chain Mobility functions}
\label{sec:mobili}

The single-chain mobility function relates the thermodynamic driving
force on monomers of a chain at point $\rr'$ to the resulting flux of
monomers from the same chain at point $\rr$.  Different expressions
have been proposed in the literature\cite{qi2017dynamic}. A popular
Ansatz is the so-called Debye scheme, which accounts for chain
connectivity and assumes that the entire chain moves as a unit.  It
takes $\Lambda^{(s)}_{\alpha \beta}(\rr,\rr')$ to be proportional to
the intra-chain pair correlation function $P_{\alpha \beta}(\rr,\rr')$,
and approximates the latter by the pair correlation function of an
ideal Gaussian chain. In Fourier representation, this gives
\begin{equation}
    \hat{\Lambda}^\text{Debye}_{\alpha \beta}(\qq) 
     = D \: \frac{1}{N} \: g_{\alpha\beta}(\qq,0)
    \label{Eq:mobili-debye}
\end{equation}
%FS: The factor k_B T disappeared because I included it in
%    rescaling scheme in this paper.
where $D$ is the diffusion coefficient of the chain, $N$ is the chain
length, and $g_{\alpha \beta}(\qq,t)$ is the partial dynamic single-chain
structure factor with respect to monomer species $\alpha$ and $\beta$.
The components of $g_{\alpha \beta}(\qq,t)$ characterize the
time-dependent correlation between monomers of type $\alpha$ and type
$\beta$  within a single chain:  
\begin{equation}
    g_{\alpha \beta}(\qq,t) 
     = \frac{1}{N} \sum_{m \in S_\alpha \atop n \in S_\beta} 
     \big\langle \exp (i \qq \cdot
     [\rr_m(t) - \rr_n(0)] ) \big\rangle. 
    \label{Eq:gqt}
\end{equation}
The sums $m,n$ run over the sets $S_\alpha$, $S_\beta$ of monomers 
in the chain which have type $\alpha$ and $\beta$, respectively. 
Specifically, the quantity $g_{\alpha \beta}(\qq,0) =: 
g_{\alpha \beta}(\qq)$ entering
\cref{Eq:mobili-debye} defines the static single-chain structure
factor.  In the Debye scheme, it is further approximated by the
structure factor of an ideal Gaussian chain, which can be expressed in
terms of the Debye function 
\begin{equation}
\label{Eq:Debye}
  F(x,f) = (2/x^2)(\exp(-fx) + fx -1) 
\end{equation}
with $ x = (q \Rg)^2$.
For homopolymers one gets $g(q)/N = F(x,1)$, and for linear diblock 
copolymers with A-fraction $f$, one obtains $g_\text{AA}(q)/N = F(x,f)$, 
$g_\text{BB}(q)/N = F(x,1-f)$,  and
$g_\text{AB}(q)/N = (F(x,1)- F(x,f) - F(x,1-f))/2$.  

Recently, our group proposed to calculate the mobility function from
the $\qq$-dependent mean relaxation times of $g_{\alpha \beta}(\qq,t)$
in a homogeneous reference system. In matrix notation, the resulting
expression reads\cite{mantha2020bottom, schmid2020dynamic}
%FS: The factor k_B T disappeared because I included it in
%    rescaling scheme in this paper.
\begin{equation}
\llhat^{(s)}(\qq) = \frac{1}{N^2}  \: 
\bg(\qq,0) \: \bG^{-1}(\qq) \: \bg(\qq,0),
\label{Eq:mobili}
\end{equation}
where $\llhat^{(s)}$ and $\bg$ are matrices with components
$\lhat^{(s)}_{\alpha \beta}(\qq)$ and $g_{\alpha \beta}(\qq,t)$, and
the matrix $\bG(\qq)$ is given by the integral $\bG(\qq) =
\frac{q^2}{N} \int_0^\infty \bg({q},t)\ud t$.

In isotropic systems such as the ones considered here, all quantities
are independent of the direction of $\qq$ and will therefore often be
expressed as a function of $q = |\qq|$. Furthermore\rdel{.}\rev{,} we define the
total single-chain dynamic structure of the chain s $g(\qq,t) =
\sum_{\alpha, \beta } g_{\alpha \beta}(\qq,t)$, and the total
single-chain mobility as $\lhat^{(s)} = \sum_{\alpha \beta}
\lhat^{(s)}_{\alpha \beta}(\qq)$.

In previous work, we have calculated the mobility functions for
unentangled Rouse chains, as defined in \cref{Eq:mobili}, both
analytically and computationally through Brownian dynamics
simulations~\cite{schmid2020dynamic}.  In the following, we extend
this work to entangled chains: On the one hand, we analytically
calculate the mobility functions for highly entangled chains using the
reptation model. On the other hand, we compute the mobility functions
from molecular dynamics simulations for a wide range of chain lengths,
spanning the unentangled to the moderately entangled regime. These
simulations were performed using the fully-flexible standard
Kremer-Grest bead-spring model~\cite{kremer1990dynamics}: All beads
have mass $m$ and interact via the purely repulsive
Weeks-Chandler-Anders\rdel{o}\rev{e}n potential\cite{WCA}, characterized by length
scale $\sigma$ and energy scale $\varepsilon$.  The bond-stretching
interactions along a chain are described by a FENE potential which
prevents chain crossings.  All quantities are expressed in units of
$\sigma$, $\varepsilon$, and $m$, i.e., the basic time unit is $\tau =
\sqrt{m \sigma^2/\varepsilon}$.  The simulations were carried out at
temperature $k_\text{B}T = 1\varepsilon$, FENE spring constant $k
= 30 \varepsilon/\sigma^2$ and bead density $\rho = 0.85
\sigma^{-3}$. For this system, we have previously~\cite{behbahani2024relaxation} estimated the statistical segment size to
be $b=1.32 \sigma$, the Kuhn length $l_\text{k}=1.81 \sigma$,   and the entanglement length \rev{associated with the plateau modulus to be
$N_\text{e}=52$~\cite{likhtman2007linear,behbahani2024relaxation, behbahani2026stress}. This value of $N_\text{e}$ gives a a tube diameter of $a = N_\text{e}^{1/2}b = 9.5 \sigma$. It should be noted that somewhat larger values of $N_\text{e}$, e.g., $78$ and $87$, have also been reported in the literature based on primitive-path analysis~\cite{moreira2015direct,everaers2020kremer}. }
We considered chains of lengths of $N = 30, 100,
200, 400$, and $N =1000$ in simulation boxes containing 800, 400,
400, 192, and 216 chains, respectively.  Therefore, the range of
chain length simulated here spans from the unentangled to the
moderately entangled regime. The monomeric friction coefficient of the
employed bead-spring model, as determined for long chains,
is~\cite{kremer1990dynamics} $\zeta = 25\tau^{-1}$. In the following,
we use this value for the expression of the mobility functions in
units of $(N\zeta)^{-1}$. We used the LAMMPS simulation
package~\cite{thompson2022lammps} with the Langevin thermostat
and a time step of $\ud t = 0.01 \tau$.

\begin{figure}[!htb]
    \centering
        \includegraphics[width=0.35\textwidth]{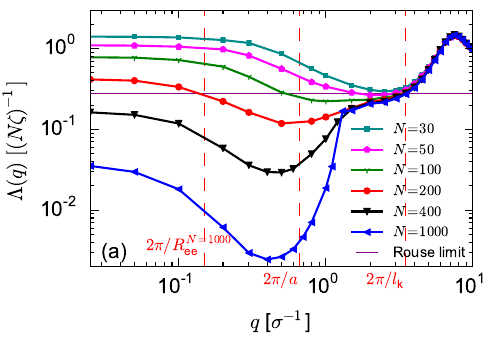}
        \includegraphics[width=0.35\textwidth]{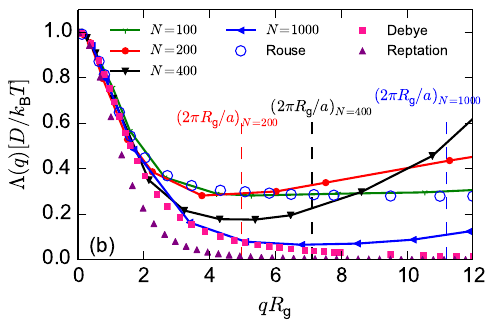}
        \includegraphics[width=0.35\textwidth]{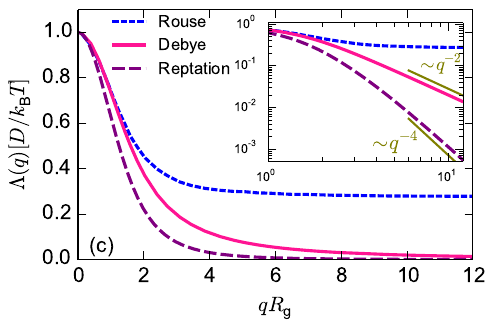}
        \caption{
        Simulation results and theoretical predictions for
        non-local mobility functions $\Lambda(q)$ of homopolymer chains
        vs. wavevector $q$.
        (a) Simulation results $\Lambda(q)$ in units of $[(N\zeta)^{-1}]$,
        for chains of different lengths as indicated.
        Error bars estimated from block averaging are comparable to the symbol size.
        The dashed vertical lines indicate $q$-values corresponding
        to the (chain-length independent) Kuhn length $l_\text{k}$ and
        tube diameter $a$, and to the end-to-end distance $R_\text{ee}$ 
        of the longest chains ($N=1000$).  
        (b) Same data for chain lengths $N \ge 100$ in units of
        $[D/\kBT]$ plotted against $q \Rg$ (solid lines)
        compared with predictions of the Rouse model 
        \cite{schmid2020dynamic} (open blue circles), 
        the Debye model \cref{Eq:mobili-debye} (red squares),
        and the reptation model (calculated from \cref{Eq:esc-rep}, 
        purple triangles).
        (c) Predictions of the Rouse model, the reptation model, 
        and the Debye scheme for the non-local mobility function 
        $\lhat(q)$, in linear (main panel) and
        logarithmic representation(inset). 
        }
    \label{Fig:mobili-homo} 
\end{figure}

In the following subsections, we present and discuss the single-chain
mobility functions obtained for homopolymers and symmetric diblock
copolymer, which we will later use in DSCFT studies of polymer blends
and copolymer melts. All mobility functions were calculated from the
average relaxation time of single-chain dynamic structure factors in
homogeneous melts, where all interaction parameters were equal as
described above, resulting in a vanishing effective Flory Huggins
parameters $\chi$.

\subsection{Homopolymers}

For homopolymers, the matrix $\bg(\qq,t)$ has only one component, the
total dynamic structure factor of the chains, and \cref{Eq:mobili}
simplifies to
\begin{equation}
 \lhat(q) = \frac{g(q,0)}{q^2 N \tau(q)} 
 \label{Eq:mobili-homo}
\end{equation}
with $\tau(q) = \int_0^{\infty} [g(q,t)/g(q,0)] \ud t $
being the mean relaxation time of the total single-chain
dynamic structure factor at wave vector $\qq$.

\cref{Fig:mobili-homo} shows the simulation results for the non-local
mobility function $\lhat({q})$ of homopolymer chains of various
lengths, calculated using the above equation.  
\cref{Fig:mobili-homo}a) presents $\lhat({q})$ in units of
$[(N\zeta)^{-1}]$ where $\zeta$ is the monomeric friction coefficient.
According to the Rouse model, the mean relaxation time of $g(q,t)$ at length
scales much smaller than the chain size ($q \gg 2\pi/R_\text{ee}$) is
chain length independent and scales as $\tau(q) \sim q^{-4}$.
Furthermore, at these length scales, the structure factor of a
Gaussian chain scales as $g(q)/N \sim q^{-2}$. Therefore,
\cref{Eq:mobili-homo} predicts that $\lhat({q})$ is constant
for $q \gg 2\pi/R_\text{ee}$.  The value of the constant was
previously calculated to be $0.279$~\cite{schmid2020dynamic}, which is
consistent with the simulation data shown in \cref{Fig:mobili-homo}a)
down to the Kuhn scale ($q \sim 2 \pi/l_\text{k}$). On even smaller 
length scales, $q > 2\pi/l_\text{k}$, the local chain stiffness becomes
important and the scaling $g(q) \sim q^{-2}$ is no longer valid
\cite{behbahani2024relaxation}, the behavior of $\lhat({q})$ deviates 
from the Rouse limit. In the high-$q$ regime, the mobility function 
is largely independent of the chain length. At lower $q$, qualitative
differences between unentangled and entangled chains become apparent. 
For short unentangled chains, the mobility increases monotonically
with decreasing $q$ until it reaches the $q \to 0$ limit. For
entangled chains ($N \gg N_\text{e} = 52$), entanglement effects 
become significant as $q$ approaches the scale of the inverse 
tube diameter.  This leads to an increase in $\tau(q)$ and, consequently, 
to a decrease in $\lhat({q})$.

\cref{Fig:mobili-homo}b) shows $\lhat({q})$ in the unit of
$[D/k_\text{B}T]$  as a  function of $qR_\text{g}$. In this
representation, which uses the natural units of DSCFT, the small-$q$
behavior of $\lhat({q})$ is more clearly resolved.  At length scales
larger than the chain size, $q \ll 2\pi/R_\text{ee}$, $g(q,t)$ is
controlled by the diffusion coefficient of the chains, $g(q,t)/g(q,0)
\approx  \exp(-Dq^2 t)$ and $\tau(q) \approx 1/(Dq^2)$.  In this
limit, \cref{Eq:mobili-homo} gives $\lhat({q}) \approx D \:
g(q)/(k_\text{B}TN)$ for homopolymers, which is also predicted in the
Debye scheme, \cref{Eq:mobili-debye}.  \cref{Fig:mobili-homo}b) also
includes the prediction of the Rouse model (blue circles). The
simulation result for $\lhat(q)$ at a chain length of $N = 100$, which
is only about twice the entanglement length $N_\text{e}$, agrees
closely with this theoretical prediction.  For longer chains ($N >
100$), entanglement effects lead to a gradual decrease in $\lhat({q})$
within the range  $2\pi/R_\text{ee} < q < 2\pi/a$ as the chain length
increases.  A more detailed discussion of the behavior of $\lhat(q)$
in entangled chains requires an extended analysis of the
time-dependence of $g(q,t)$; for homopolymers, such an analysis was
provided in our previous work~\cite{behbahani2024relaxation}.

For highly entangled chains, we can derive an analytical estimate for
$\lhat({q})$ from the tube model.  According to that model, the
behavior of $g(q,t)$ for entangled chains is governed by two
processes: (i) the fast process of local movement of the chain inside
the tube, and (ii) the slow escape of the chain from its confining
tube~\cite{de1981coherent}.  At length scales much larger than the
tube diameter ($q \ll 2\pi/a$), which are relevant for DSCFT
calculations for very long chains, the fast process can be neglected.
At such length scales, reptation dynamics predicts the following
expression for $g(q,t)$ of highly entangled
chains~\cite{doi1988theory}:
\rev{\begin{equation}
        g(q,t) 
        = \sum_{p=1}^{\infty}\left[\frac{2\mu N}
          {\alpha_p ^2(\mu^2 + \alpha_p^2 + \mu)}
   \sin^2{\alpha_p} \exp(-\frac{4 D_\mathrm{c} t \alpha_p^2}{L^2})
   \right],
\label{Eq:esc-rep}
\end{equation}
Here, $L = Za$ is the contour length of the tube, $Z = N/N_\mathrm{e}$ is the number of entanglement strands per chain (or the number of steps of the tube), and $a$ is the tube step length (or tube diameter). The quantity $D_\mathrm{c} = k_\mathrm{B}T/(N\zeta)$ is the curvilinear diffusion coefficient of the chain along the tube, which is related to the real-space diffusion coefficient by $D = D_\mathrm{c}/(3Z)$. }  The parameter $\mu = q^2R_\mathrm{ee}^2/12$, and the $\alpha_p$ are the positive solutions of $\alpha_p\tan\alpha_p=\mu$.
\rev{It should be noted that the above equation assumes pure reptation as the sole mechanism for chain escape from the tube. It therefore neglects contour-length fluctuations (CLF), which govern chain dynamics up to the Rouse time, $\tau_\text{R}$~\cite{mcleish2002tube,wischnewski2002molecular,behbahani2024relaxation}. This approximation is consistent with our main goal of calculating mobility functions for very long, highly entangled chains, as the separation between $\tau_\text{R}$ and the disengagement time, $\tau_\text{d}$, increases with chain length. In that regime, the relaxation time $\tau(q)$ obtained from the integral of the single-chain dynamic structure factor $g(q,t)$ is therefore increasingly dominated by pure reptation.
CLF also modifies the magnitude and the chain-length scaling of the disentanglement time, relative to pure reptation. However,  since we derive the mobility function in terms of the diffusion coefficient (as discussed below), this modification is implicitly incorporated through the chain-length dependence of $D$.}

\rev{The argument of the exponential in the \cref{Eq:esc-rep} can equivalently be expressed as $(-12Dt\alpha_p^2/R_\mathrm{ee}^2)$.} The above expression can \rev{then}
be integrated straightforwardly to obtain $\tau(q)$ as a function of
$D$.  The resulting $\tau(q)$, together with the Debye expression for
$g(q)$, can then be inserted into \cref{Eq:mobili-homo} to obtain
$\lhat({q})$ in units of $[D/(k_\text{B}T)]$ as a function of
$qR_\text{g}$. In this representation, the resulting mobility function
becomes chain-length independent, as shown in
\cref{Fig:mobili-homo}b-c). 

For homopolymers, the mobility $\lhat({q})$ estimated from the Debye
scheme lies between the mobilities predicted by the Rouse and
reptation models.  The $q$ dependence of  $\lhat({q})$ in the regime
$q \gg 2\pi/R_\text{ee}$ is particularly noteworthy: in this range,
the Rouse model predicts a constant mobility ($\sim q^0$), the Debye
scheme yields $\lhat(q) \sim q^{-2}$, and the reptation model predicts
$\lhat(q) \sim q^{-4}$.  This difference between the Rouse and the
reptation models originates from the qualitatively different dynamics
of unentangled and entangled chains. In the regime $q \gg
2\pi/R_\text{ee}$, the relaxation time of Rouse chains scales as
$\tau(q) \sim q^{-4}$, whereas for entangled chains, $\tau(q)$ becomes
independent of $q$ and is equal to disentanglement
time~\cite{doi1988theory}. This difference in the scaling behaviors of
$\tau(q)$ is reflected in the scaling behavior of $\lhat({q})$.   

\subsection{Diblock copolymers}

We proceed with the calculation of the single-chain mobility matrix
for symmetric AB diblock copolymers. Due to the symmetry, it has only 
two independent components, $\lhat_\textrm{AA}({q})$
($=\lhat_\textrm{BB}({q})$) and $\lhat_\textrm{AB}({q})$
($=\lhat_\textrm{BA}({q})$), and the same holds for the matrices
$\bg$ and $\bG$. Therefore, diagonalizing all matrices, one
can decompose \cref{Eq:mobili} as follows~\cite{mantha2020bottom}:
\begin{align}
    \begin{split}
        \lhat_\textrm{AA} ({q}) &= \frac{1}{4q^2N}
          \left( \frac{g(q,0)} {\tau(q)} 
              + \frac{\Delta(q,0)}{\tau_\Delta({q})} \right)\\
        \lhat_\textrm{AB} ({q}) &= \frac{1}{4q^2N}
           \left( \frac{g({q},0)} {\tau({q})} 
              - \frac{\Delta({q},0)}{\tau_\Delta({q})} \right).
    \end{split}
    \label{Eq:mobili-copolym}
\end{align}
Here $\tau({q})$ is the wave vector-dependent mean relaxation time of
the overall single-chain dynamic structure factor $\tau(q) =
\int_0^\infty \ud t \: g(q,t)/g(q,0)$.  The function
$\Delta({q},t) = g_\textrm{AA}({q},t) + g_\text{BB}({q},t) -
g_\textrm{AB}({q},t) - g_\textrm{BA}({q},t)$ quantifies the difference
between like and unlike correlations, and its associated relaxation
time is defined as $\tau_\Delta = \int_0^\infty \ud t
\Delta({q},t)/\Delta({q},0)$. 

\begin{figure}[tb]
    \centering
        \includegraphics[width=0.35\textwidth]{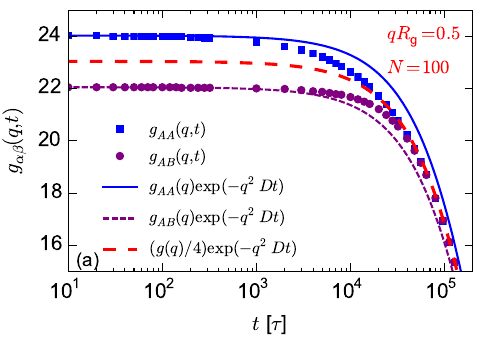}
        \includegraphics[width=0.35\textwidth]{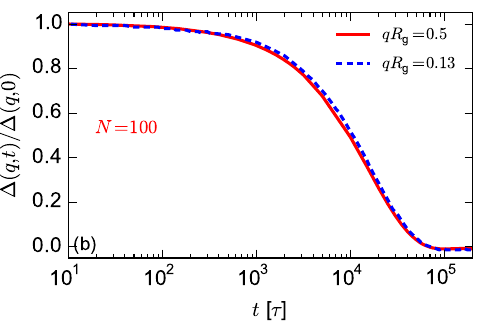}
        \includegraphics[width=0.35\textwidth]{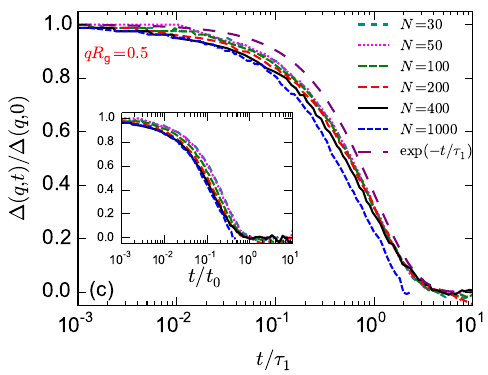}
  \caption{Characterization of partial single-chain dynamic structure factors
    $g_{\alpha \beta}(q,t)$ of symmetric diblock copolymer
    chains in a dense disordered melt ($\chi=0$). (a) Simulation results
    (symbols) for $g_\textrm{AA}(q,t)$ and $g_\textrm{AB}(q,t)$ at
    copolymer chain length $N=100$ at $q \Rg=0.5$, 
    compared to different approximations as indicated (lines). 
    (b) Rescaled function $\Delta(q,t)/\Delta(q,0)$ with
    $\Delta(q,t) = 2(g_\textrm{AA}(q,t) - g_\textrm{AB}(q,t))$ 
    versus time $t$ for copolymers with length $N=100$ 
    at two wave vectors $q$, both in the regime $q < 2\pi/\Ree$. 
    (c) Same function $\Delta(q,t)/\Delta(q,0)$ plotted against  
    $t/\tau_1$, where $\tau_1$ is the longest relaxation time of the 
    end-to-end vector autocorrelation function, for various chain lengths $N$ 
    as indicated and fixed $q\Rg = 0.5$. The inset displays 
    $\Delta(q,t)/\Delta(q,0)$ as a function of $t/\tO$ with
    $\tO = \Rg^2/D$, the natural time unit in DSCFT calculations.
    \label{Fig:delta} }
\end{figure}

To understand the behavior of the mobility matrix, we
examine the behavior of $\bg(q,t)$ in the limit of
small $q$ and late times. In the Supporting Information (SI),
section S1, we analyze the asymptotic behavior of $\bg(q,t)$ for
general heteropolymers with arbitrary sequences and
architectures, and show that it has the following
general form both in the limits $t \to \infty$ and
$t=0$ (Eq.\ (S9)):
\begin{equation}
    \frac{1}{N}g_{\alpha \beta}(\qq,t)  \approx
    \ue^{- D q^2 t} \: 
    \left( I_\alpha(\qq) I_\beta(\qq)
      + \frac{q^2}{3} c_{\alpha \beta}(t) 
   + {\cal O}(q^4) \right)
 \label{eq:gqt_asym}
 \end{equation}
 \begin{align}
  \nonumber
  \textrm{with} \qquad &
  I_\alpha(\qq) =  \frac{1}{N}
  \big\langle \sum_{n \in S_\alpha} \:
  \ue^{\ui \, \qq \cdot \uu_n} \big \rangle
  \approx
  f_\alpha \: 
    \ue^{- \frac{q^2}{6} \langle \uu^2 \rangle_\alpha}
 \\
\textrm{and} \qquad &
   c_{\alpha \beta}(t) =
    \frac{1}{N^2}
     \sum_{n \in S_\alpha \atop m \in S_\beta}
        \langle \uu_n(t) \cdot \uu_m(0) \rangle.
  \nonumber
  \end{align}
Here $\uu_n(t)=\rr_n(t) - \RR_\textrm{c}(t)$ denotes the distance
of a monomer $n$ to the geometric center of the molecule, 
$\RR_\text{c}(t)= \frac{1}{N} \sum_n \rr_n(t)$
(implying $\langle \uu_n \rangle = 0$), and $\langle \uu^2
\rangle_\alpha$ is the average squared distance of monomers of type
$\alpha$.  The first term in \cref{eq:gqt_asym} describes the dominant
contribution of the slow translational diffusion mode. The other terms
subsume the effect of the rotational diffusion and higher order
internal movements of the chains, which decay to zero on the 
characteristic time scale $\tau_1$ of the chain relaxation.

Specifically for symmetric copolymers, \cref{eq:gqt_asym} gives the 
following asymptotic behaviors for wave vectors
$q \ll \sqrt{1/(\tau_1 D)}$ (\ie, setting $\ue^{-Dq^2 \tau_1} \approx 1$):
\begin{align}
\label{eq:gt-asymp}
\begin{split}
g(q,t) &= N \ue^{- D q^2 t} + {\cal O}(q^4) \\
\Delta(q,t) &= N \frac{2}{3} q^2 \:
\Delta c(t)
+ {\cal O}(q^4) ,
\end{split}
\end{align}
with $\Delta c(t) = \big(c_\textrm{AA}(t) - c_\textrm{AB}(t)\big)$,
corresponding to
\begin{equation}
g_\textrm{AA/AB} (q,t) \approx 
 \frac{N}{4} \: \ue^{- D q^2 t} 
 \pm \frac{N}{6} \: q^2 \:
\Delta c(t).
\label{eq:g-asymp}
\end{equation}
This behavior can be explicitly seen in \cref{Fig:delta}a) which shows
$g_\textrm{AA}(q,t)$ and $g_\textrm{AB}(q,t)$ at $q\Rg = 0.5$ for a
symmetric diblock chain with length $N = 100$. At late times,
$g_{\alpha \beta}(q,t)$ approaches $g_\textrm{AA}(q,t) =
g_\textrm{AB}(q,t) = g(q,t)/4 = g(q)/4\exp(-q^2 Dt)$.
\cref{Fig:delta}b) shows $\Delta(q,t)/\Delta(q,0)$ for two different
values of $q$, $qR_\text{g} = 0.5$ and $qR_\text{g} = 0.13$, both
corresponding to length scales larger than the chain size. The results
show that  $\Delta(q,t)/\Delta(q,0)$ is independent of $q$ in the
range $q \ll 2\pi/R_\text{ee}$, as expected from
\cref{eq:gt-asymp}. \rev{Physically, this means that the relaxation 
of internal modes on short wavelengths is dominated by a single process,
the relaxation of average monomer displacements $\uu_n(t)$ relative to 
the center of mass (averaged over monomer species). For symmetric copolymers, 
this becomes apparent in the time-dependent behavior of $\Delta(q,t)$. For asymmetric
copolymers, the decomposition of the mobility matrix into
diffusive and internal contributions is more complicated and discussed
in detail in SI, section SI-1.}

The characteristic time scale of $\Delta(q,t)$ in the small-$q$ regime
is most clearly illustrated in \cref{Fig:delta}c), which presents
$\Delta(q,t)/\Delta(q,0)$ as a function of $t/\tau_1$ for various
chain lengths $N$ at fixed $q\Rg = 0.5$.  For each chain length,
$\tau_1$ was calculated as the longest relaxation time of the
end-to-end vector autocorrelation
function~\cite{behbahani2024relaxation}.  The curves closely follow
$\exp(-t/\tau_1)$ and nearly collapse when plotted against $t/\tau_1$.
The inset of \cref{Fig:delta}c) instead graphs
$\Delta(q,t)/\Delta(q,0)$ versus $t/\tO$, where $\tO = R_\text{g}^2/D$
is the natural unit of time in DSCFT calculations. In both the Rouse
and the reptation models, $\tO$ is proportional to the longest chain
relaxation time, $\tO = (\pi^2/2)\tau_1$. One would therefore expect
that plotting against $t/\tO$ instead of $t/\tau_1$ simply shifts the curves
horizontally on the logarithmic axis. However, \rev{we find that
the ratio $\tau_1/\tO$ exceeds the theoretical value for 
small chain lengths, $N < 100$, (e.g., $\tau_1/\tO \sim 1.3 \cdot 2/\pi^2$ at $N=30$), reflecting non-ideal behavior of short chains in polymer melts\cite{meyer2008static}. As a consequence,} the curves \rev{therefore} \rdel{also} become slightly 
more spread out. To test whether \rdel{this is due to the
system-size dependence of}\rev{the influence of finite size effects on} 
the diffusion coefficient $D$, and thus \rdel{of }$\tO = R_\text{g}^2/D$, 
we performed additional test simulations
with varying box sizes. With the range of sizes considered here, the
resulting variation with $\tO$ was small compared to the differences
observed in the inset of \cref{Fig:delta}c. 

Since \cref{eq:gt-asymp} is valid both for large $t$ and at $t = 0$
we can use it to analyze the structure of the mobility functions
at $q\to 0$.  The mean-relaxation time is then dominated by the late-time 
behavior, and \cref{Eq:mobili-copolym} yields
\begin{equation}
\lhat_\textrm{AA/AB}(q) =
\frac{1}{4} \left( D \pm \: \frac{1}{3} 
   \frac{\Delta c(t)}{\int_0^\infty \ud t' \Delta c(t')}
   \right) + {\cal O}(q^2) 
\label{eq:mobility-asymp}
\end{equation}
The first term represents the overall diffusion of the chains, and the
second term reflects rotational and other internal motions. Due to
the contributions of the latter, the components of $\llhat$ 
differ from each other even at $q = 0$.

We proceed with a discussion of the predictions of the reptation model
for $\Delta(q,t)$ and the single-chain mobility functions of symmetric
diblock copolymers. For highly entangled linear chains, Doi and
Edwards derive an analytical expression for the correlation function
$\langle \exp\{\ui \, \qq \cdot [\rr(s,t) - \rr(s',0)] \} \rangle$ as
a function of the contour distance $s$ from a chain end in their
classic textbook~\cite{doi1988theory}.  Using this expression, one can
straightforwardly calculate $g_\textrm{AA}(q,t)$,
$g_\textrm{AB}(q,t)$, and $\Delta(q,t)$ for symmetric diblock
copolymers in a homogeneous melt. The details of the calculations and
the results for $g_\text{AA}(q,t)$ and $g_\text{AB}(q,t)$ are provided
in SI, section SI-2. For $\Delta(q,t)$ of a reptating symmetric diblock
copolymer, the result is 
\begin{align}
    \begin{split}
\Delta(q,t) &= \sum_{p=1}^{\infty}[\frac{2\mu N}{\beta_p ^2(\mu^2 + \beta_p^2 + \mu)}
  \\ & \quad \times 
     (1-\cos \beta_p)^2 \exp(-\frac{12 t D \beta_p^2}{R_\text{ee}^2})],
    \end{split}
    \label{Eq:delta-rep}
\end{align}
with $\mu = q^2 R_\text{ee}^2/12$  and $\beta_p$ being the solutions
of $\beta_p \cot \beta_p = - \mu$.  Specifically in the $q$-range
$q\ll 2\pi/R_\text{ee}$, one has $\mu \ll 1$ and the above expression is
dominated by the first term containing $\beta \approx \pi/2$.
Consequently $\Delta(q,t) \approx (16 q^2 R_\text{g}^2  N/\pi^4)
\exp(-t/\tau_1)$, where $\tau_1 = (2/\pi^2) R_\text{g}^2/D$ is the
longest relaxation time of the chain.  Therefore, the relaxation time
of $\Delta(q,t)$ becomes independent of $q$ for $q\ll
2\pi/R_\text{ee}$, consistent with our earlier considerations and
with the simulation results shown in \cref{Fig:delta}.

\begin{figure}[tb]
    \centering
        \includegraphics[width=0.35\textwidth]{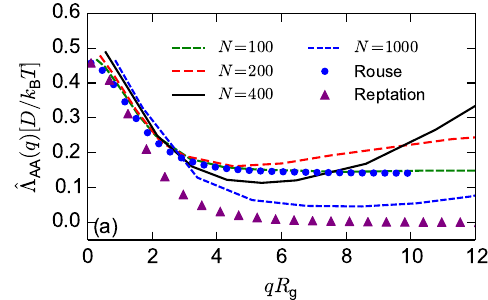}
        \includegraphics[width=0.35\textwidth]{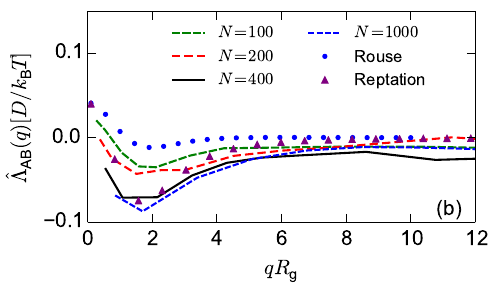}
        \includegraphics[width=0.35\textwidth]{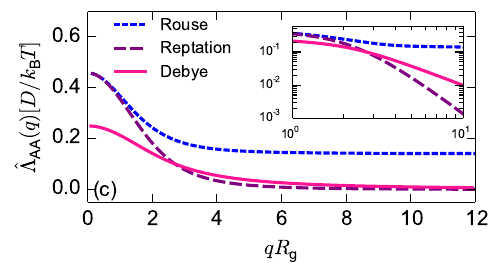}
        \includegraphics[width=0.35\textwidth]{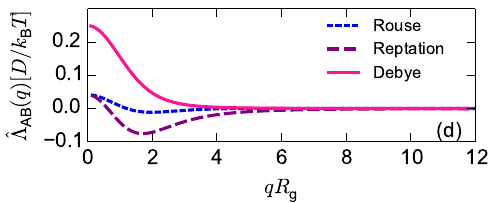}
        \caption{Scaled mobility functions $\lhat_{\alpha \beta}(q)$ of 
        symmetric diblock copolymers.
        (a,b): Simulation results from simulations of disordered melts
        for varying chain lengths (lines), compared with predictions 
        from the Rouse model (blue circles) and the \rdel{repration}\rev{reptation} model
        (purple triangle), for $\lhat_\textrm{AA}$ (a) and
        $\lhat_\textrm{AB}$ (b). 
        (c,d): Corresponding theoretical predictions from the Rouse model 
        (short dashed blue line), the reptation model (long dashed
        purple line), and the Debye scheme (pink line).
        The inset in panel (c) shows the data on a logarithmic scale.   
    \label{Fig:mobili-copolym}} 
\end{figure}

Based on these predictions (\cref{Eq:esc-rep} and \cref{Eq:delta-rep}) and
using \cref{Eq:mobili-copolym}), we compute the single-chain mobility
functions for symmetric diblock copolymers in the highly entangled
regime.  The results are presented in \cref{Fig:mobili-copolym} along
with the theoretical prediction for unentangled Rouse
chains\cite{schmid2020dynamic}, as well as with simulation data for
chains of varying lengths.  For comparison, we also include the mobility 
functions of the Debye scheme, $\lhat_{\alpha \beta}^\text{Debye}(q)$
(\cref{Eq:mobili-debye}).
In the small-$q$ range, the mobility
functions $\lhat_{\alpha \beta}(q)$ computed from $\bg(q,t)$ differ
qualitatively from the Debye-based functions $\lhat_{\alpha
\beta}^\text{Debye}(q)$.  For $q \rightarrow 0$, both
$\lhat_\text{AA}^\text{Debye}(q)$ and
$\lhat_\text{AB}^\text{Debye}(q)$ approach $0.25$, whereas the
Rouse and reptation models predict $\lhat_\text{AA}(0) \approx
0.458$ and $\lhat_\text{AB}(0) \approx 0.0416$.  As discussed 
above, this difference arises from the contributions of
rotational and internal chain modes to the components of the
mobility functions.

The chain-length dependence of $\lhat_\text{AA}(q)$ in
\cref{Fig:mobili-copolym} resembles the behavior of $\lhat(q)$ for a
homopolymer chain shown in \cref{Fig:mobili-homo}.  For $N = 100$,
$\lhat_\text{AA}(q)$ closely follows the prediction of the Rouse
model. For longer chains ($N > 100$), $\lhat_\text{AA}(q)$ gradually
decreases in the range of $2\pi /R_\text{ee} < q < 2\pi/a$ and
approaches the prediction of the reptation model. The behavior of
$\lhat_\text{AB}(q)$ differs from that of
$\lhat_\text{AA}(q)$. For $q > 2\pi/R_g$, $\lhat_\text{AB}(q)$ is
nearly zero, while for $q < 2\pi/R_g$, it exhibits a minimum with a
negative value.  This minimum becomes more pronounced for entangled
chains. At $q \to 0$, the limiting behavior of $\lhat_{\alpha
\beta}(q)$ as obtained from simulations for longer chains slightly
deviates from the predictions of the Rouse and reptation models.  This
deviation is consistent with the spread observed in the
$\Delta(q,t)/\Delta(q,0)$ curves for different chain lengths when
plotted against $t/\tO$, (inset of \cref{Fig:delta}c).

\section{Comparison between DSCFT calculations and MD simulations}
\label{sec:md}

Having determined the single-chain mobility functions, we proceed to
assessing the accuracy of the resulting DSCFT models. To this end, we
compare DSCFT predictions for the spinodal decomposition in symmetric
binary homopolymer blends and symmetric diblock copolymer melts
with corresponding molecular dynamics (MD) simulation data.
We first focus on deterministic DSCFT models. The influence of
adding stochastic noise will be discussed in the next section.

The MD simulations of inhomogeneous systems were performed using the
same model as before, the fully flexible Kremer-Grest model, with one
difference: In order to make monomers A and B immiscible, the
prefactors of the Weeks-Chandler-Anderson non-bonded potentials
between like and unlike monomers differ from each other. Following
previous studies~\cite{grest1996efficient,murat1999statics}, we set
$\varepsilon_\text{AA} = \varepsilon_\text{BB} = \varepsilon$ and
$\varepsilon_\text{AB} / \varepsilon_\text{AA} = 1 + e$, where the
parameter $e > 0$ controls the strength of the relative repulsion
between A and B monomers.  

We consider polymers with chain lengths $N = 100$ and $N = 400$.  The
corresponding invariant degrees of polymerization, determined from the
end-to-end distance $\Ree$ {\em via} $\bar{N} = (\rho \Ree^3/N)^2$,
are given by $337$ and $1517$, respectively. Since $R_\text{ee}$
exhibits slight deviations from Gaussian scaling in this range of $N$,
the ratio of $\bar{N}$ values is larger than the corresponding ratio
of chain lengths $N$. The Ginzburg parameters of the systems
are $G=1.33$ for $N=100$, and $G=2.66$ for $N=400$.

Chains with $N = 100$ are only twice the entanglement length
($N_\text{e} = 52$) and exhibit Rouse-like behavior.  In contrast,
chains with $N = 400$  contain approximately eight entanglement
strands and show signatures of entangled dynamics. This difference in
the dynamics is reflected in their mobility (see
\cref{Fig:mobili-homo} and \cref{Fig:mobili-copolym}). For $N= 100$,
$\lhat(q)$ closely follows the Rouse prediction, whereas for $N =
400$, $\lhat(q)$ shows significant deviations indicating entanglement
effects.  

In the MD simulations of structure formation, a cubic simulation box
with side length $6.6 \Rg$ was used, corresponding to 360 and 768
chains for $N=100$ and $N=400$, respectively. The systems are
initially prepared in a homogeneous state at $e = 0$, after which $e$
is suddenly increased to a value above the demixing transition (for
blends) or the order-disorder transition (for copolymer melts). The
configurations are then analyzed at selected times to monitor the time
evolution of the system.  To improve the statistics, 20 independent
simulations were performed with different initial configurations  for
each parameter set.

The DSCFT calculations employed a box of same size than the MD
simulations, $V = (6.6 \Rg)^3$.  It was discretized with a grid
spacing of $0.3 \Rg$, resulting in 22 grid points per spatial
dimension, and periodic boundary conditions were applied in all
directions. The SCFT propagator equation was solved using a
pseudo-spectral method~\cite{muller2005incorporating} combined with
Anderson mixing~\cite{anderson1965iterative}, and the polymer contour
was discretized into 100 segments. The time step in the DSCFT
calculations was $10^{-6} \tO$, where $\tO = \Rg^2/D$. The initial
density fields were generated by mapping the corresponding initial MD
configurations onto the grid using a combination of a cloud-in-cell
(CIC) mapping scheme~\cite{qi2017dynamic} and a Gaussian filter. In
the CIC scheme, each bead contributes to the densities of eight
surrounding grid points with weights inversely proportional to the
distance. 

To quantitatively compare the structure of MD and DSCFT
configurations, we inspect the compositional structure factor $S(\qq)$
at $\qq \neq 0$. In the MD simulations, it is calculated from
\begin{equation}
    S(\qq) 
= \frac{1}{N_\textrm{beads}} \bigg| \sum_j 
\Psi_j \exp\left( \ui \, \qq \cdot \rr_j \right) \bigg|^2,
\label{Eq:sq_md}
\end{equation}
where the sum runs over all beads in the system, $N_\text{beads}$
is the total number of beads, 
and $\Psi_j = \pm 1/2$ depending on the type of monomer $j$ 
($+1/2$ for A and $-1/2$ for B). Defining the compositional density 
field $\Psi(\rr) = \sum_j \Psi_j \delta(\rr - \rr_j)$, this expression 
can be rewritten as
\begin{equation}
    S(\qq) 
   = \frac{1}{N_\textrm{beads}} \Big| \Psi(\qq) \Big|^2.
\label{Eq:sq}
\end{equation}
In grid calculations, where scaled concentrations 
of monomer species are specified on a set of grid points,
we calculate a corresponding quantity based on 
$\psi(\rr) = \Psi(\rr) / \rho_0 =
(\phi_\textrm{A}(\rr)- \phi_\textrm{B}(\rr))/2$,
\begin{equation}
   \Sgrid(\qq) = \frac{1}{N_\text{grid}} \bigg| \sum_k \psi(\rr_k)
       \exp\left( \ui \, \qq \cdot \rr_k \right) \bigg|^2,
\label{Eq:sq_grid}
\end{equation}
where the sum now runs over $N_\textrm{grid}$ grid points $k$ at
positions $\rr_k$. Comparing \cref{Eq:sq} and \cref{Eq:sq_grid} and
using $\rho_0 = N_\textrm{beads}/V$ and $\int_V \ud^3 r \: f(\rr)
\approx \sum_k v_\textrm{g} \: f(\rr_k)$ with the voxel volume
$v_\textrm{g} = V/N_\textrm{grid}$, one can easily see that
$\Sgrid(\qq)$ has to be multiplied with the factor $\rho_0 v_g
= {N_\textrm{beads}}/{N_\textrm{grid}}$ to match $S(\qq)$.  In a finite
periodic simulation box, $S(q)$ can only be calculated for a discrete set
of $\qq$-vectors that are commensurate with the box dimensions: $q_i =
2\pi n_i/l_i$ with $i \in \{ x,y,z\}$, $l_i$ being the box length in the
$i$-direction, and $n_i$ an integer.  We average over all $\qq$-vectors
with equal magnitude and express the structure factor as a function of
$q = |\qq|$.

\subsection{Matching the effective interaction parameter}

For a quantitative comparison of results from MD simulations and DSCFT
calculations, the relative repulsion parameter $e$ ($=
\varepsilon_\textrm{AB}/\varepsilon_\textrm{AA} - 1$) must be mapped
onto a corresponding Flory-Huggins parameter $\chi$.  Following Morse
and Chung~\cite{morse2009chain}, we use the following estimate:
\begin{equation}
    \chi(e) \approx \frac{e z_\infty}{k_\text{B}T} \quad
    \text{with} \quad
    z_\infty = \lim_{N\to \infty} \: z(N)
    \label{Eq:chi}
\end{equation}
where $z(N) = \int g_N^\textrm{inter}(r) \: u(r) \: 4\pi r^2 \: \ud r
$ is the effective coordination number for chains of length $N$.  Here
$g_N^\textrm{inter}(r)$ denotes the intermolecular radial pair
distribution function for a homopolymer melt composed of chains of
length $N$, and $u(r)$ is the non-bonded pair potential energy.  To
calculate $z_\infty$, we extrapolated $z(N)$ values obtained from
simulations of homopolymer melts with varying chain lengths, up to $N
= 1000$, yielding $z_\infty = 56.7$ and thus $\chi(e) = 56.7 e$. The
corresponding simulation data and calculation details are presented in
SI, Section SI-3 and Fig.\ S1. In the older
literature\cite{muller1995computer,murat1999statics} Flory Huggins
parameters were estimated using expressions similar to \cref{Eq:chi},
however using $z(N)$ instead of $z_\infty$. For the chain lengths
studied in the present work ($N = 100$ and $N = 400$), the estimates
based on $z(N)$ are close to those based on $z_\infty$.

\begin{figure}[tb]
    \centering
        \includegraphics[width=0.4\textwidth]{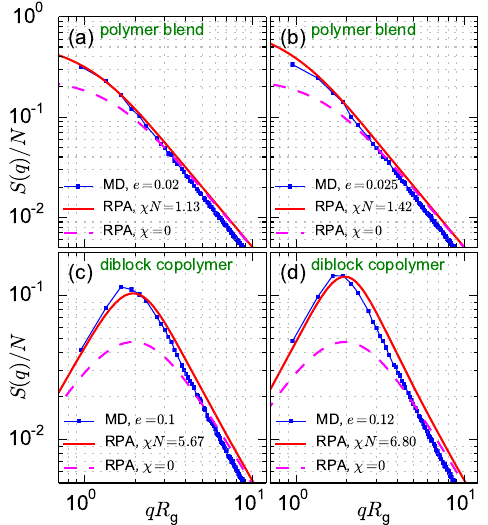}
           \caption{Compositional structure factors $S(q)$ 
           in the disordered state from MD simulations 
           with different relative repulsion parameters $e$ 
           (blue line with markers) compared with RPA predictions at
           $\chi=0$ (magenta dashed line) and $\chi(e)=59.7 e$ 
           (see \protect\cref{Eq:chi}, red solid line). 
           (a,b) Binary homopolymer melts; (c,d) diblock copolymer
           melts
    \label{Fig:Sq-RPA}} 
\end{figure}

To assess the reliability of the estimate $\chi(e) = 56.7e$, we
simulated disordered phases of polymer blends and diblock copolymer
melts with chain length $N = 100$ at selected values of $e$ below the
order-disorder transition. We then compared the structure factors
obtained from the MD simulations with the predictions of the random
phase approximation (RPA) for $\chi = 56.7 e$ It should be noted that
the RPA is exact only in the limit of very long
chains~\cite{qin2012fluctuations}. For inhomogeneous melts composed of
chains of finite length, more refined theories are required to
accurately describe the structure
factor~\cite{grzywacz2007renormalization,qin2011renormalized,
qin2012fluctuations,glaser2014collective}.  However, the limitations
of the RPA are most pronounced near the order-disorder transition. Far
from the transition -- for example at $\chi N \approx 5$ in copolymer
melts -- the RPA provides a reasonable
approximation~\cite{qin2012fluctuations}.

In the random phase approximation (RPA) for incompressible polymer
melts, the compositional structure factor is given by 
$S(q) = N\ \Gamma_2^{-1}(q)$ with ~\cite{schmid2011}:
\begin{equation}
\Gamma_2 = \left( \frac{K_\text{AA} + K_\text{BB} 
  + K_\text{AB} + K_\text{BA}}{K_\text{AA} K_\text{BB} 
  - K_\text{AB} K_\text{BA}} - 2 \chi N \right),
\label{Eq:RPA}
\end{equation}
where $K_{\alpha \beta}(\qq) = \sum_{(\gamma)} \bar{\phi}^{(\gamma)}
\:g_{\alpha\beta}^{(\gamma)}(q)/N$ is the sum over scaled single-chain
structure factors of (co)polymers of type $\gamma$, weighted with
their global volume fraction $\bar{\phi}^\gamma$.  For copolymer
melts, this simply gives $K_{\alpha \beta} = g_{\alpha \beta}(q)/N$.
For binary homopolymer blends with average fraction $\bar{\phi}_A$ of
$A$-segments, one gets $K_\text{AA} = \bar{\phi}_A \,
g_\text{AA}(q)/N$, $K_\text{BB} = (1-\bar{\phi}_A) \,
g_\text{BB}(q)/N$, and $K_\text{AB} = 0$. The function $g_{\alpha
\beta}$ is taken to be equal to partial structure factor of an ideal
Gaussian chain, which can be expressed in terms of the Debye function,
see paragraph following \cref{Eq:Debye}. 

\begin{figure*}[tb]
    \centering
        \includegraphics[width=0.7\textwidth]{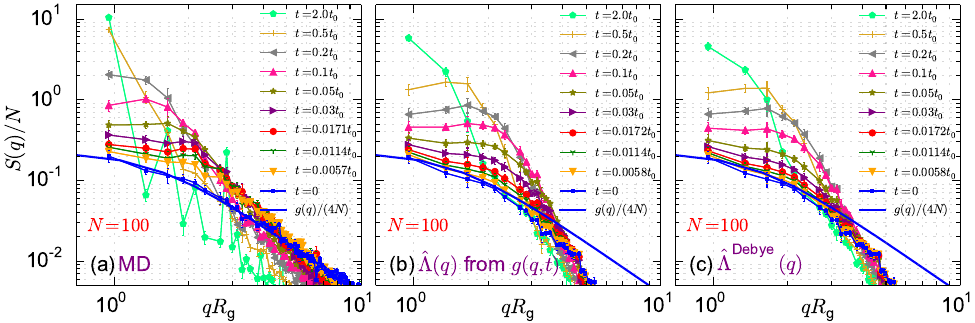}
        \includegraphics[width=0.7\textwidth]{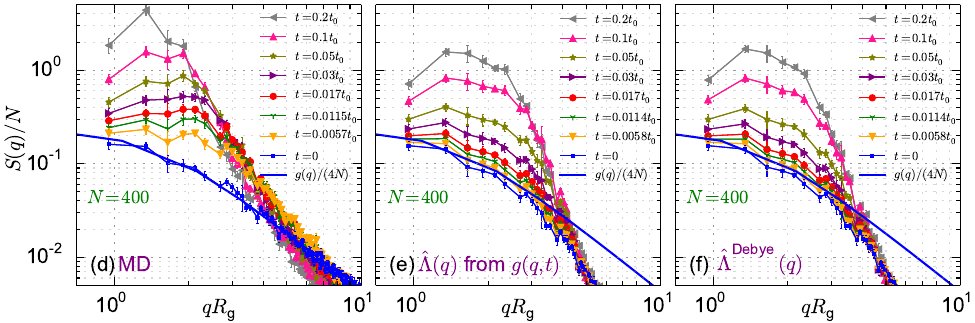}
        \includegraphics[width=0.7\textwidth]{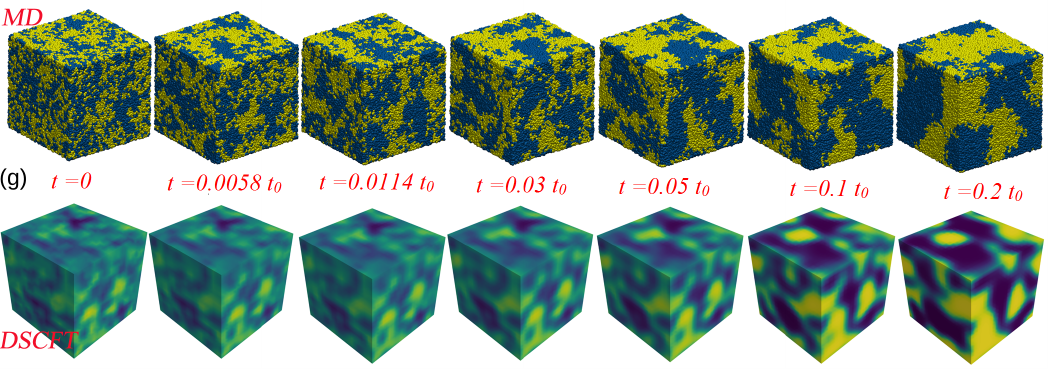}
        \caption{Dynamics of spinodal decomposition in symmetric
        polymer blends following a quench from $\chi N = 0$ to $\chi N
        = 28.35$. Panels (a–c) show the reduced structure factor
        $S(q)/N$ at different times (in units of $t_0 =
        R_\text{g}^2/D$) for melts with chain length $N = 100$, and
        panels (d–f) show results for $N = 400$, as obtained from MD
        simulations (a, d), DSCFT with mobility from \rev{MD simulation data for}
        $g(q,t)$ (b, e),
        and DSCFT with the Debye mobility (c, f). Panel (g) shows
        system snapshots for $N = 400$ at different times from MD
        simulations (top) and DSCFT with mobility derived from 
        $g(q,t)$ (bottom).}
        \label{Fig:Sq-blend} 
\end{figure*}

\cref{Fig:Sq-RPA} shows the structure factor $S(q)$ obtained from MD
simulations for the disordered phases of polymer blends and diblock
copolymer melts with chain length $N = 100$ at various values of the
repulsion parameter $e$ between unlike monomers. The figure also shows
the corresponding RPA predictions with the Flory-Huggins parameter
$\chi(e) = 59.7 e$ derived from \cref{Eq:chi}, and those for $\chi =
0$ for reference. Overall, good agreement is observed between the
calculated $S(q)$ curves and the RPA predictions, confirming the
reliability of the relation $\chi(e) = 56.7 e$ within the examined
range of $e$. At large $q$, the RPA prediction slightly deviates from
the simulation results. \rdel{These}\rev{This} discrepancy can be attributed to
deviations of the chain conformations from ideal Gaussian statistics.
This source of error can be removed by directly computing the reduced
single-chain structure factor $K_{\alpha\beta}$ in \cref{Eq:RPA} from
the simulations, rather than approximating it using Debye functions
(see \cref{Fig:Sq-blend}a,d) and \cref{Fig:Sq-copolymer}a,d), 
blue lines and symbols).

Knowing the relation between the parameters controlling the
incompatibility of A and B monomers in the MD and DSCFT models, $e$
and $\chi$, we are now ready to compare MD simulations with DSCFT
computations. The remaining unknown parameter of the DSCFT model is the
inverse compressibility $\kappa$, which we set to $\kappa N = 100$ to
describe nearly incompressible fluids.

\begin{figure}
    \centering
     \includegraphics[width=0.4\textwidth]{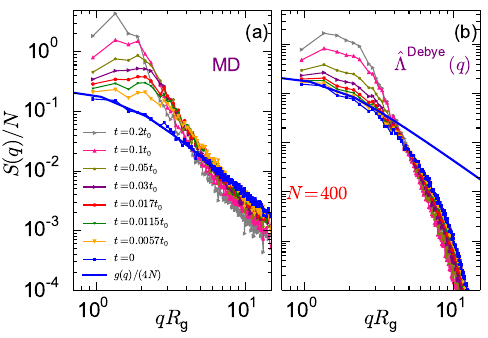}
     \caption{Reduced structure factor, $S(q)/N$, at different
     times for blends with chain length $N = 400$. Panels (a) and (b)
     show the results of MD simulations and DSFCT with Debye mobility,
     respectively. Compared to \cref{Fig:Sq-blend}, the data are shown
     over a wider $q$-range, which emphasizes the high-$q$
     region.} \label{Fig:Sq-blend-highq} 

\end{figure}

\begin{figure*}[tb]
    \centering
        \includegraphics[width=0.7\textwidth]{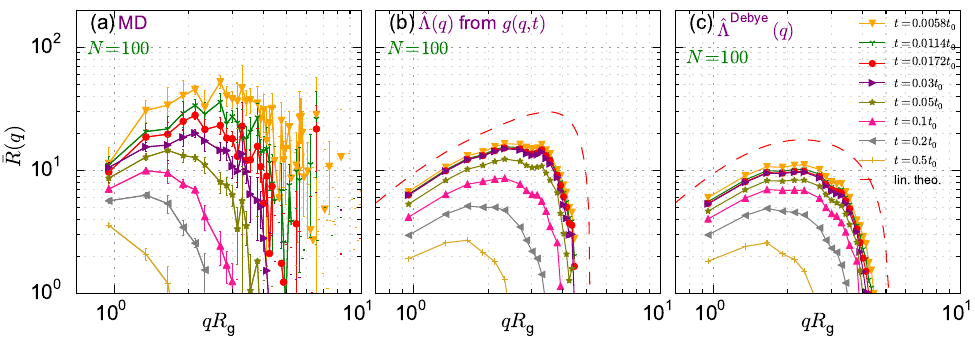}
        \includegraphics[width=0.7\textwidth]{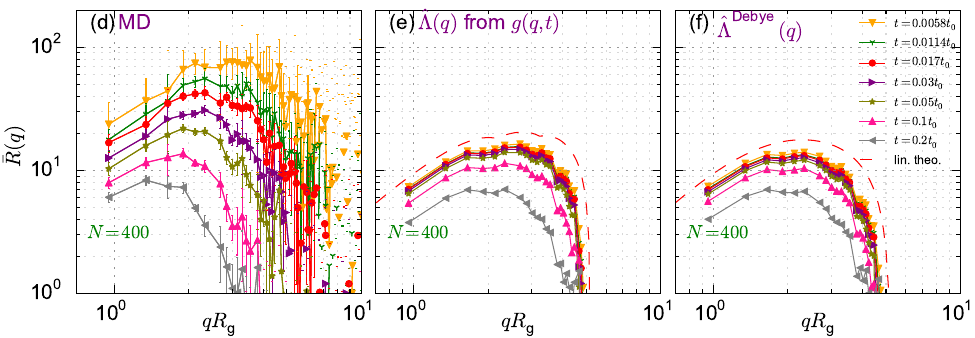}
        \caption{Average growth rate of the structure factor,
        $\bar{R}(q)$, at different times in symmetric polymer blends
        following a quench from $\chi N=0$ to $\chi N = 28.35$. Panels
        (a–c) show $\bar{R}(q)$ for blends with chain length $N =
        100$, (d–f) for $N = 400$. Results are from MD simulations (a, d), 
        DSCFT with mobility from \rev{MD simulation data for} $g(q,t)$ 
        (b,e), and DSCFT with the Debye 
        mobility (c,f). The dashed red lines show the predictions of 
        the linearized theory (\cref{Eq:Rt}). The error bars on the MD data 
        represent the standard error of the mean obtained from $20$
        independent runs. The error bars on the DSCFT data are
        comparable to the marker size.   }
    \label{Fig:Rq-blend} 
\end{figure*}

\begin{figure} 
    \centering
        \includegraphics[width=0.35\textwidth]{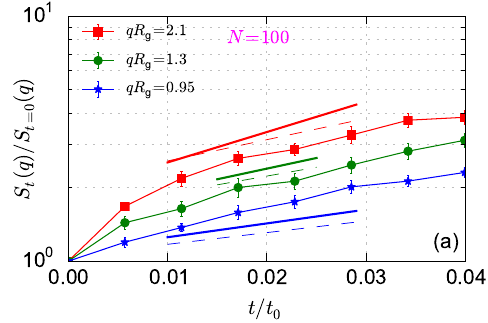}
        \includegraphics[width=0.35\textwidth]{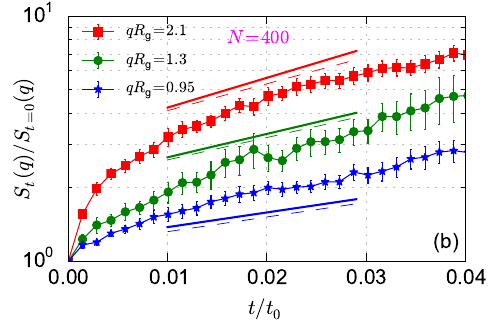}
        \caption{Time evolution of the normalized structure factor,
        ($S_t(q)/S_{t=0}(q)$), following a quench from ($\chi N=0$) to
        ($\chi N=28.35$) from MD simulations of symmetric blends with
        chain length $N=100$ (a) and $N=400$ (b). Results are shown
        for three wave vectors $q$ over the time interval where DSCFT
        predicts approximately exponential  growth (cf.
        \cref{Fig:Rq-blend}). Solid lines indicate the corresponding 
        DSCFT growth rates for $\bg(q,t)$-based mobilities\rev{, 
        while thin dashed lines show the growth slopes obtained 
        using the Debye mobility. }
        }
    \label{Fig:Ratio-blend} 
\end{figure}
\subsection{Demixing of homopolymer blends}

We first consider the dynamics of spinodal decomposition in symmetric
polymer blends, starting from an initially homogeneous mixture
prepared at $e=0$, i.e., $\chi N = 0$.  \cref{Fig:Sq-blend} g)
compares representative configuration snapshots from MD simulations
($N=400$) and DSCFT computations at different times following a quench
to $e=0.125$, corresponding to $\chi N = 28.35$. Visually, the time
evolution in the two models seems very similar. 

For a more quantitative comparison, we examine the evolution of the
reduced structure factor $S(q)/N$, shown in \cref{Fig:Sq-blend}
(a-c) for chain length $N=100$ and in \cref{Fig:Sq-blend}(d-f) for chain
length $N=400$. Panels (a,d) display the MD results, (b,e) the 
SCFT results using the single-chain mobility derived from the
single-chain dynamic structure factor according to \cref{Eq:mobili},
and (c,f) the DSCFT results obtained with the Debye mobility,
\cref{Eq:mobili-debye}.  In the MD simulations, the relative repulsion
parameter was set to $e=0.5$ for $N=100$, and to $e=0.125$ for
$N=400$, yielding $\chi N = 28.35$ in both cases according to
\cref{Eq:chi}. 

According to the RPA prediction, \cref{Eq:RPA}, the structure factor
at $\chi N = 0$ should reduce to $S(q) = g(q)/4$ for a symmetric
blend, where $g(q)$ is the single-chain structure factor. This
relation is clearly confirmed by the MD results shown in
\cref{Fig:Sq-blend}(a,d) for the initial configurations (time $t=0$,
blue), which were prepared as equilibrated blends at $\chi N=0$.  The
data also demonstrate the accuracy of the RPA when $g(q)$ is computed
directly from the simulations rather than approximated by the Debye
function as in \cref{Fig:Sq-RPA}.  

In contrast, in the DSCFT results shown in
\cref{Fig:Sq-blend}b,c,e,f), $S(q)$ at $t=0$ deviates from $g(q)/4$ at
high $q$. The discrepancy can be attributed to the mapping procedure used
for generating initial configuration for the DSCFT calculations from the
MD data. The mapping modifies the structure at small length scales,
leading to distortions in the high-$q$ tail of $S(q)$.  In
\cref{sec:noise} we will show how to avoid this artifact by including
stochastic fluctuations in the DSCFT equation.

Next we compare the MD and DSCFT predictions for the time evolution of
$S(q)$.  Following a quench to $\chi N = 28.35$ into the two-phase
region, long-wavelength concentration fluctuations grow spontaneously,
leading to an increase of $S(q)$ at low $q$ in both MD and DSCFT. The
structure factor develops the characteristic peak of spinodal
decomposition at nonzero $q$, reflecting the spinodal pattern
formation.  Subsequently, the peak gradually increases in magnitude
and shifts to smaller $q$, indicating progressive coarsening of the
structure. The growth of $S(q)$ is generally faster in the MD
simulations, than in the DSCFT calculations, particularly at very
short times. Apart from that, the DSCFT calculations follow the MD
simulations reasonably well.  In the high-$q$ regime, however,
differences emerge, as also shown more clearly in \cref{Fig:Sq-blend-highq}.
In the MD simulations, $S(q)$ decreases slightly during phase
separation, but remains close to $g(q)/4$ until the late stages. In
contrast, DSCFT calculations yield significantly lower values of
$S(q)$ in this regime, indicating a strong suppression of
short-wavelength fluctuations. The DSCFT curves feature a crossover
wave vector $q^*$ at which $S(q)$ remains approximately
time-independent at short times, i.e., $S_t(q^*) =
S_\text{t=0}(q^*)$).  For $q<q^*$, $S(q)$ increases, whereas for $q >
q^*$, it decreases. Such a crossover is not observed in the MD
results. As will be shown later, these discrepancies arise from the
neglect of stochastic fluctuations in the DSCFT calculations.

The two DSCFT approaches give \rdel{very} similar results. This reflects the
fact that, in the $q$-range relevant for structure formation, the
mobilities of chains with $N=100$ and $N=400$ remain close to the
Debye mobility (see \cref{Fig:mobili-homo}b).  Small systematic
differences can be traced to differences in the shape of the mobility
functions.  As shown in \cref{Fig:mobili-homo}, for $q >
2\pi/R_\text{ee}$, the mobilities derived from $\bg(q,t)$ exceed
the Debye mobilities for both chain lengths,
with the deviation being more pronounced for $N = 100$.
Consequently, DSCFT calculations employing the $\bg(q,t)$-based mobility
(panels b and e of \cref{Fig:Sq-blend}) exhibit a slightly faster growth
of $S(q)$ than those using the Debye mobility (panels c and f),
particularly at short times. 
\rev{To facilitate the comparison between the MD simulation results 
and the DSCFT results obtained using different choices of mobility 
functions, Fig. S2 in SI shows the $S(q)/N$ curves at a fixed time 
($t = 0.03 t_0$) for each chain length, with the results from the 
different methods plotted together in a single panel. The 
$S(q)/N$ curves obtained using the $g(q,t)$-based mobility 
functions are slightly higher and closer to MD results than those 
obtained using the Debye mobility, particularly for $N = 100$ 
and at intermediate $q$ values. } 
\rdel{These differences gradually diminish
as the system evolves.}

For completeness, we note that we have also performed DSCFT
calculations without enforcing proportionality between the flux and
the local chain density, i.e., replacing the local chain density in
\cref{Eq:flux-local-rho}) with the global average
density\cite{qi2017dynamic}. While this formulation reproduced the
same short-time behavior of $S(q)$, it became numerically unstable
at later stages of phase separation.

Comparing systems of chains with different lengths, one observes 
a more pronounced spinodal peak in the structure factor (at $q \Rg
\sim 2$) for $N=400$ than for $N=100$, both in MD simulations and
DSCFT calculations. Moreover, the peak persists for longer times at
$N=400$.  One might be tempted to attribute this behavior to
differences in the mobility functions. However, the same trend is
observed in DSCFT calculations employing the Debye mobility, which
does not distinguish between entangled and unentangled dynamics. We
therefore conclude that the effect does not originate from
entanglements, but rather attribute it to the ($1/N$) scaling
of density fluctuations in the initial configurations, which affects
the importance of nonlinear contributions. In systems of longer chains,
the initial fluctuations are smaller, allowing spinodal patterns to grow 
to larger amplitudes  before nonlinear coarsening effects become significant.

More information on the dynamics of spinodal decomposition can be
obtained by analyzing the average growth rate of the structure factor,
defined as~\cite{binder2001spinodal}:
\begin{equation}
    \bar{R}(q) = \frac{\ln(S_t(q)/S_{t=0}(q))}{2 t}. 
    \label{Eq:Rt}
\end{equation}
Time-independence of $\bar{R}(q)$ implies exponential evolution of
$S(q)$, i.e., $S(q) = S_{t=0}(q) \exp(2 \bar{R}(q) t )$. Such a
behavior is predicted within the RPA theory, which assumes that
the fluctuations of the local density are sufficiently small
that the free energy functional in \cref{eq:energy-scf} 
entering \cref{Eq:DDFT-phi-deterministic} can be linearized. 
For a symmetric polymer blend, one obtains~\cite{binder2001spinodal} 
(see also SI, Section \rdel{S4}\rev{SI-5})
\begin{equation}
        \bar{R}^\mathrm{blend}(q) 
         = - \frac{1}{4} q^2 \lhat(q) \Gamma_2(q), 
\label{Eq:Rt-linear-homo}
\end{equation}
where $\lhat(q)$ is the single-chain mobility function and
$\Gamma_2(q)$ is obtained from \cref{Eq:RPA}. In the regime where
$\Gamma_2(q) < 0$, the static structure factor $S(q)$ grows, whereas
for $\Gamma_2(q) > 0$, it decays. For a symmetric blend, 
$\Gamma_2(q) = 4.0 / F(x,1) - 2\chi N$, where $F(x,1)$ is
the Debye function introduced in \cref{Eq:Debye}.

\begin{figure*}[!htb]
    \centering
        \includegraphics[width=0.7\textwidth]{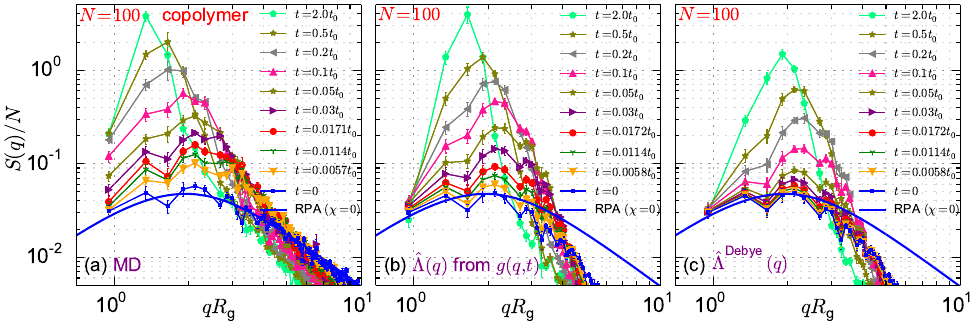}
        \includegraphics[width=0.7\textwidth]{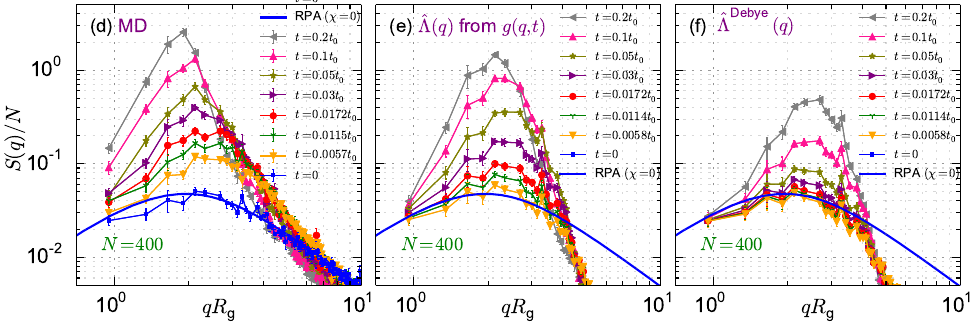}
        \includegraphics[width=0.7\textwidth]{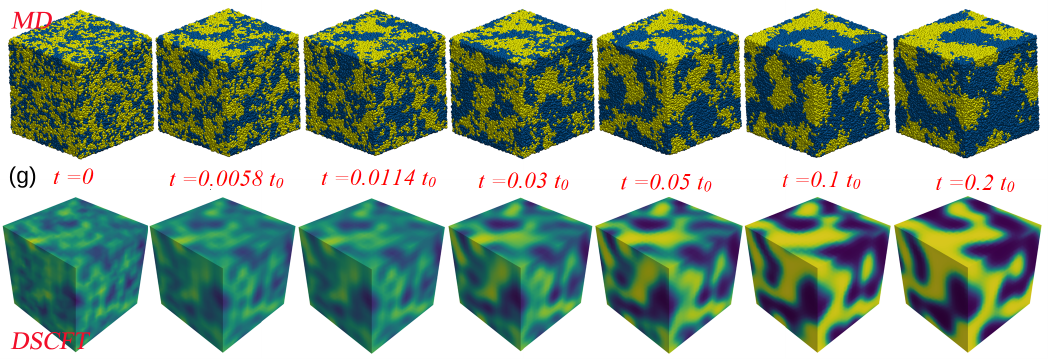}
     \caption{Dynamics of microphase separation in diblock copolymer
     melts after a quench to $\chi N = 28.35$. Panels (a–c) show the
     reduced structure factor $S(q)/N$ at different times for melts
     with chain length $N = 100$, while panels (d–f) show results for
     $N = 400$. Results are from MD simulations (a, d), DSCFT with
     mobility from \rev{MD simulation data for} $\bg(q,t)$ (b, e), 
     and DSCFT with the Debye mobility (c, f). Panel (g) shows system 
     snapshots for $N = 400$ at
     different times from MD simulations and DSCFT with mobility
     derived from $\bg(q,t)$.  }
    \label{Fig:Sq-copolymer} 
\end{figure*}

\begin{figure*}[!htb]
    \centering
        \includegraphics[width=0.7\textwidth]{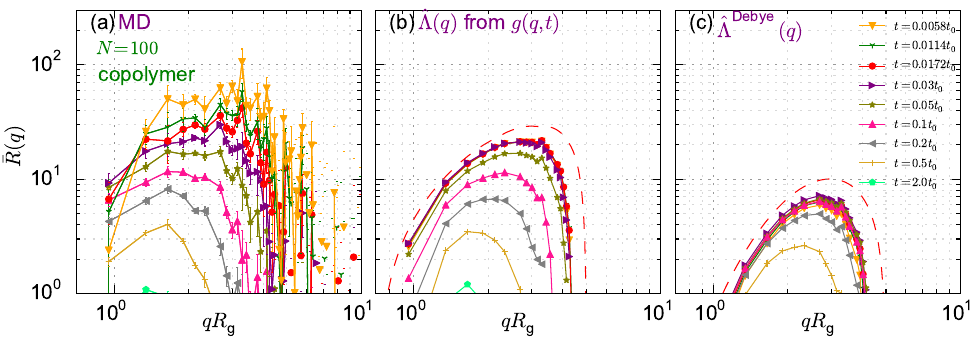}
        \includegraphics[width=0.7\textwidth]{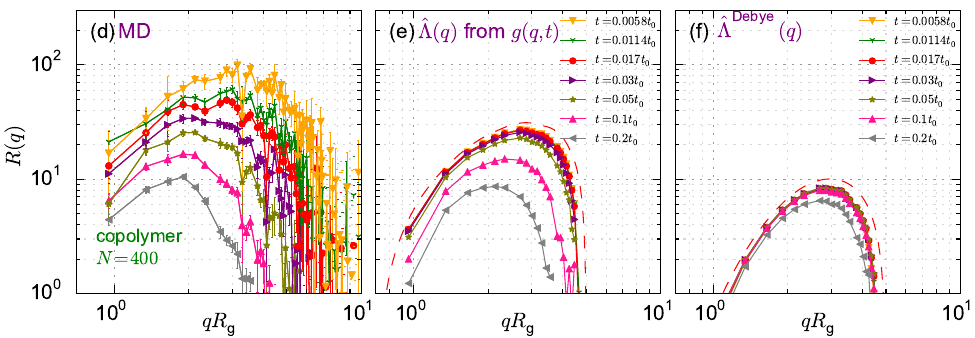}
     \caption{Average growth rate $\bar{R}(q)$ at different times
     during microphase separation in copolymer melts after a sudden
     quench to $\chi N = 28.35$.  Panels (a–c) show $\bar{R}(q)$ fo
     chain length $N = 100$, and panels (d–f) for $N = 400$. 
     Results are from MD simulations (a, d), DSCFT with mobility
     from \rev{MD simulation data for} $\bg(q,t)$ (b, e), and DSCFT with the Debye mobility (c, f).
     The dashed red lines represent the predictions of the linearized
     theory (\cref{Eq:Rt}). The error bars on the MD data were
     calculated as the standard error of the mean from $20$
     independent runs. The error bars on the DSCFT data are comparable
     to the marker size. } 
     \label{Fig:Rq-copolymer} 
\end{figure*}
\begin{figure}
    \centering
        \includegraphics[width=0.35\textwidth]{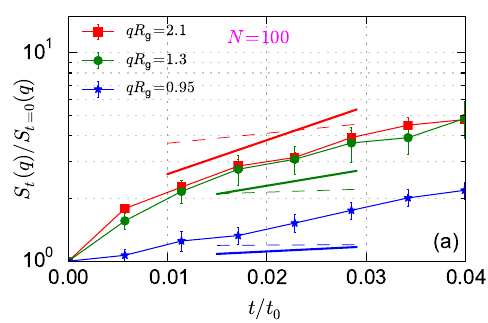}
        \includegraphics[width=0.35\textwidth]{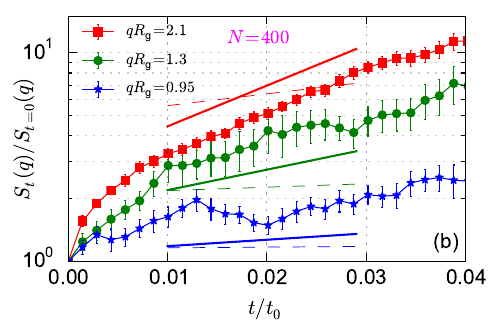}
        \caption{
        Time evolution of the normalized structure factor, ($S_t(q)/S_{t=0}(q)$), 
        following a quench from ($\chi N=0$) to ($\chi N=28.35$) from MD simulations of 
        symmetric diblock copolymers with chain length $N=100$ (a) and $N=400$ (b)
        for three wave vectors $q$. The wave vector $q\Rg=2.1$ is approximately
        at the structure-factor peak. Solid lines indicate the exponential
        growth rates predicted by DSCFT with $\bg(q,t)$-based mobilities in the
        same time interval\rev{, while thin dashed lines show the growth rates 
        obtained using the Debye mobility.}}
        \label{Fig:Ratio-copolymer} 
\end{figure}

\cref{Fig:Rq-blend} shows the average growth rates calculated from MD
simulations and DSCFT calculations for blends with chain lengths of $N
= 100$ and $N = 400$. As in \cref{Fig:Sq-blend}, the DSCFT results are
presented for both the mobility derived from $\bg(q,t)$ and the Debye
mobility, together with the prediction of the linearized RPA theory
\cref{Eq:Rt-linear-homo}. Within DSCFT, the growth rate $\bar{R}(q)$
remains nearly constant at early times, consistent with an exponential
growth of $S(q)$. At later times, $\bar{R}(q)$ decreases, and the
range of $q$ over which it is positive shrinks. The linearized theory
overestimates the initial growth rate, but the agreement with DSCFT
improves for the longer chains, where the fluctuations in the initial
density field are weaker.

In contrast, the average growth rate measured in MD is initially large 
and decreases with time. Since $\bar{R}(q)$ is averaged over a finite 
time interval, this can mask variations in the instantaneous growth rate. 
To examine these variations, \cref{Fig:Ratio-blend} shows the time 
evolution of $S_t(q)/S_{t=0}(q)$ on a semi-logarithmic scale, where 
exponential growth appears as a straight line. Results are presented for 
blends with $N=100$ and $N=400$ and three wave vectors $q$, up to the time
$0.04 \tO \approx 0.20 \tau_1)$. In this time period, DSCFT predicts 
approximately exponential growth of the structure factor (see 
\cref{Fig:Rq-blend}), with growth rates as indicated by the solid
lines \rdel{(}for $\bg(q,t)$-based mobilities\rdel{)}\rev{, 
and with dashed lines for the Debye mobility}. The MD results reveal 
a rapid initial increase of $S_t(q)/S_{t=0}(q)$, particularly near the 
peak wave vector \rev{at $q \Rg \sim 2$}, followed by a regime of approximately exponential growth.
The growth rates in this second regime are \rev {generally} in good agreement with the DSCFT predictions.
\rev{For the examined $q$ values, the growth rates predicted by the $g(q,t)$-based and Debye mobilities are close to each other, particularly for the two smaller $q$ values and for the longer chains with $N = 400$. This is consistent with the comparable $S(q)$ curves obtained using the two choices of mobility functions, as shown in \cref{Fig:Sq-blend} and Fig. S2 in SI. At $qR_\text{g} = 2.1$, the growth rate predicted by the Debye mobility is approximately $30\%$ and $10\%$ lower than that predicted using the $g(q,t)$-based mobility for chains with $N = 100$ and $N = 400$, respectively. We also fitted a line to the data of $\ln[S_t(q)/S_{t=0}(q)]$ versus $t$ obtained from MD simulations in the range $0.01 t_0$ to $0.03 t_0$. The differences between the resulting growth rates and those predicted by the DSCFT calculations are generally less than $30\%$.}

Recently, M\"{u}ller and coworkers~\cite{wang2019collective,
steffen2025collective} investigated the short-time dynamics of
spinodal decomposition in unentangled polymer melts using linearized
theory and particle-based simulations of highly coarse-grained soft
polymer chains. In their simulations, they also observed an initial
rapid increase of the structure factor, followed by a regime of
exponential growth. They showed that the initial fast increase arises from memory effects in the mobility kernels, which are neglected in
our DSCFT approach. The subsequent exponential regime extends to much
later times than in our system, up to $t > 0.1 \tO$. This can be
attributed to the lower fluctuation level in their initial
conformations. The use of a soft polymer model enabled simulations of
systems with an extremely large invariant degree of polymerization
$\bar{\cal N}=(\rho_0 \Ree^3/N)^2 = 2.56 \times 10^8$, compared to
$\bar{\mathcal N}\sim 340$ and $\bar{\mathcal N}\sim 1400$ in the MD
simulations presented here. 

\subsection{Ordering kinetics in copolymer melts}

We now turn to the dynamics of structure formation during the spinodal
decomposition in symmetric diblock copolymer melts.
\cref{Fig:Sq-copolymer} g) shows a representative series of snapshots
from MD simulations (top) and DSCFT calculations (bottom), following a
quench from a disordered state at $\chi N = 0$ to $\chi N = 28.35$,
of copolymer melts with chains length $N=400$. Like in the
blend case, visual inspection indicates very similar evolution.

The corresponding reduced structure factors, $S(q)/N$, are shown in
\cref{Fig:Sq-copolymer} (a-c) for $N=100$ and \cref{Fig:Sq-copolymer}
(d-f) for $N=400$.  For each chain length, results from from MD
simulations (a,d), DSCFT employing mobility functions derived from the
single-chain dynamic structure factor $g(q,t)$ according to
\cref{Eq:mobili} (b, e), and DSCFT using the Debye mobility (c, f),
are presented.  At $t=0$, the MD results for $S(q)$ are well described
by the RPA (\cref{Eq:RPA}). In contrast, the DSCFT results exhibit
deviations in the high-$q$ regime due to the mapping procedure used to
generate the initial continuum configurations from the MD data, as
discussed previously for homopolymer blends.  Furthermore, similar to
the blend case, DSCFT exhibits a pronounced suppression of $S(q)$  at
high $q$, reflecting the absence of stochastic fluctuations in the
present calculations.

In the low-$q$ regime, DSCFT reproduces the MD behavior; however, as
in the blend simulations, the growth of $S(q)$ is faster in MD at
early times.  Compared to blends (\cref{Fig:Sq-blend}), the
differences between the two DSCFT approaches are substantially more
pronounced for copolymers (panels (b,e) vs. panels (c,f)). 
\rev{A direct comparison of the results obtained using the 
different methods is provided in SI in Fig. S2, where the 
$S(q)/N$ curves at time $t=0.03 \tO$ from all methods are 
plotted in a single panel for both chain lengths $N=100$ and 
$N=400$.}
DSCFT calculations employing the Debye mobility predict considerably slower
dynamics than calculations using mobilities derived from $\bg(q,t)$.
This behavior can be traced to the mobility functions themselves: For
copolymers, the $\bg(q,t)$-based mobility deviates from the Debye form
both in the low-$q$ and high-$q$ regime, whereas for polymer blends
such deviations are largely restricted to the high-$q$ regime.

Next we examine the average growth rate of the compositional structure
factor, $\bar{R}(q)$ (\cref{Eq:Rt}). The RPA approximation leading
to \cref{Eq:Rt-linear-homo} for homopolymer blends can also be
applied to copolymer systems. The calculation is shown 
in SI, Section \rdel{S4}\rev{SI-5}. For incompressible (co)polymer mixtures containing 
two types of monomers, A and B, one obtains:
\begin{equation}
 \bar{R}(\qq) = - q^2 \:
   \frac{\lhat_\mathrm{AA}(\qq) \lhat_\mathrm{BB}(\qq)
       - \lhat_\mathrm{AB}(\qq) \lhat_\mathrm{BA}(\qq)}
       {\lhat_\mathrm{AA}(\qq) + \lhat_\mathrm{AB}(\qq)
        + \lhat_\mathrm{BA}(\qq) + \lhat_\mathrm{BB}(\qq) }
   \: \Gamma_2(\qq),
\end{equation}
where $\Gamma_2(\qq)$ has been defined in \cref{Eq:RPA}.
In our symmetric diblock copolymer melts, we have 
$\lhat_\mathrm{AA} = \lhat_\mathrm{BB}$,
 $\lhat_\mathrm{AB} = \lhat_\mathrm{BA}$, leading to
\begin{equation}
        \bar{R}^\mathrm{diblock}(q) = - \frac{1}{2} q^2
        [\lhat_\mathrm{AA}(q)-\lhat_\mathrm{AB}(q)] \Gamma_2(q).
\label{Eq:Rt-linear-copolymer}
 \end{equation}

\Cref{Fig:Rq-copolymer} shows the average growth rates of the structure
factor obtained from the MD and DSCFT simulations of copolymer melts
together with the predictions of the linearized RPA,
(\cref{Eq:Rt-linear-copolymer}). In DSCFT, $\bar{R}(q)$ is nearly
time-independent at early times, in good agreement with the linearized
theory, and decreases gradually at later times. For copolymers, the
early-time value of $\bar{R}(q)$ predicted using $\bg(q,t)$-based mobilities
is significantly larger than that obtained with the Debye mobility. As
in the blend case, the MD results feature a large initial growth rate
that decreases continuously during the subsequent evolution.

To clarify the instantaneous growth behavior, \cref{Fig:Ratio-copolymer} 
shows the time evolution of the normalized structure factor $S_t(q)/S_{t=0}(q)$ 
in diblock copolymer melts with chain lengths $N=100$ and $N=400$ for three
selected wave vectors, including two in the low-$q$ regime and one near 
the peak position of the structure factor, $qR_g=2.1$ (\cref{Fig:Sq-copolymer}). 
The  evolution is shown up to $0.04 \tO \approx 0.2 \tau_1$, during which 
DSCFT predicts approximately exponential growth. The slopes predicted 
by DSCFT with $\bg(q,t)$-based mobilities are indicated by solid lines 
\rev{and those predicted by the Debye mobility are shown by dashed lines}. As
in the blend case, the MD data for $S_t(q)/S_{t=0}(q)$ exhibit a rapid 
initial increase, particularly near the peak -- presumably reflecting
memory effects \cite{steffen2025collective} -- followed by approximately 
exponential growth. While the initial burst is absent in DSCFT, the 
subsequent slopes are generally in good agreement with DSCFT predictions
with $\bg(q,t)$-based mobilities. In contrast, DSCFT calculations using 
the Debye mobility predict substantially slower growth and deviate 
more significantly from the MD results. \rdel{(see Fig. 11).} 
%\rdel{(see \cref{Fig:Ratio-copolymer}).}

For the smallest $q$, both DSCFT calculations and the corresponding
linearized theories predict an almost vanishing growth rate, significantly 
smaller than that observed in MD. This is also evident 
in \cref{Fig:Sq-copolymer},  where $S(q)$ is almost constant at the 
smallest $q$ in the DSCFT calculations, whereas it grows substantially 
in the MD simulations. As we shall see in the next section, this artifact 
of DSCFT can be removed by including stochastic fluctuations in the 
DSCFT equations. 

\section{The effects of stochastic fluctuations}
\label{sec:noise}

In the framework of DSCFT, initializing the system from prescribed
(random) configurations and evolving them deterministically -- i.e.,
without incorporating stochastic fluctuations (thermal noise) -- can
lead to an incomplete exploration of the configuration space for
several reasons.  First, stochastic fluctuations are  essential for
allowing the system to escape local free energy minima and access the
full range of relevant microstructures. Second, generating
representative initial configurations is itself a nontrivial task and
can induce artifacts.  For example, the grid-based mapping
procedure used in the previous section to convert MD configurations
onto DSCFT fields imposes an unphysical cutoff at high $q$, i.e., at
short wavelengths (cf.\ blue lines in \cref{Fig:Sq-blend} and
associated discussion).

\begin{figure}
    \centering
        \includegraphics[width=0.35\textwidth]{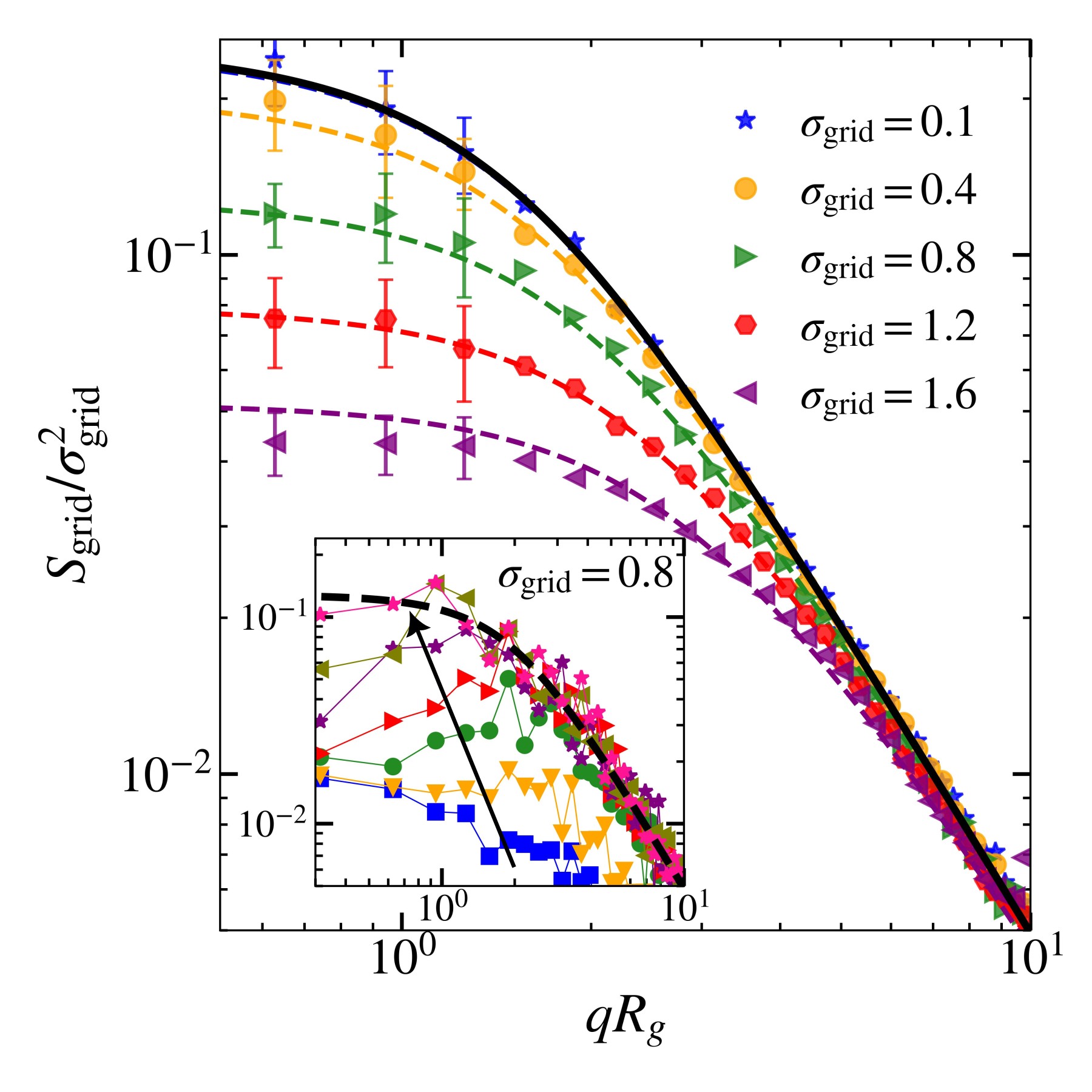}
    \caption{Reduced structure factor for different $\sgrid$ at 
    $\chi N=0$ for a homopolymer mixture. The dashed lines show
    fits to the heuristic expression $\Sgrid(q) = \ssgrid/(4+6 g(q)
    \ssgrid)$,
 black solid line the RPA prediction. The inset 
    shows the time evolution of the reduced structure factor 
    for $\sgrid=0.8$, starting from an initially fully random 
    configuration at $t=0$. 
    The curves correspond to the times
    $t/\tO =0.01, 0.03, 0.1, 0.3, 1.1, 2.1, 6.7 $ in the direction indicated by the arrow.
    }
    \label{Fig:Sq-noise-chiN0}

\end{figure}

As discussed in the section ``Dynamical self-consistent field
theory'', the standard way to incorporate thermal fluctuations in
DSCFT is to introduce stochastic currents that satisfy a
fluctuation-dissipation relation, \cref{Eq:DDFT-phi} and
\cref{Eq:noise-fluc-dis}.  In the present section, we examine the
effect of such noise terms on the DSCFT results. We first note that,
in discretized DSCFT configurations, the amplitude of density
fluctuations increases with decreasing voxel volume $v_g$, as follows
from the relation $\Sgrid(q) = \frac{1}{\rho_0 v_g} S(q)$ that
connects the grid-based and the particle-based structure factors
(Eqs.~(\ref{Eq:sq_grid}, \ref{Eq:sq})). Motivated by this scaling, we
define the grid noise amplitude $\ssgrid = \sigma^2
\cdot(\Rgcube/v_g)$, and examine the reduced grid structure factor
$\Sgrid(q)/\ssgrid$ in the following, which maps to $\frac{1}{N} S(q)$
for $\sigma^2 = 1/G$.

\begin{figure*}[t]
    \centering
   \includegraphics[width=0.9\textwidth]{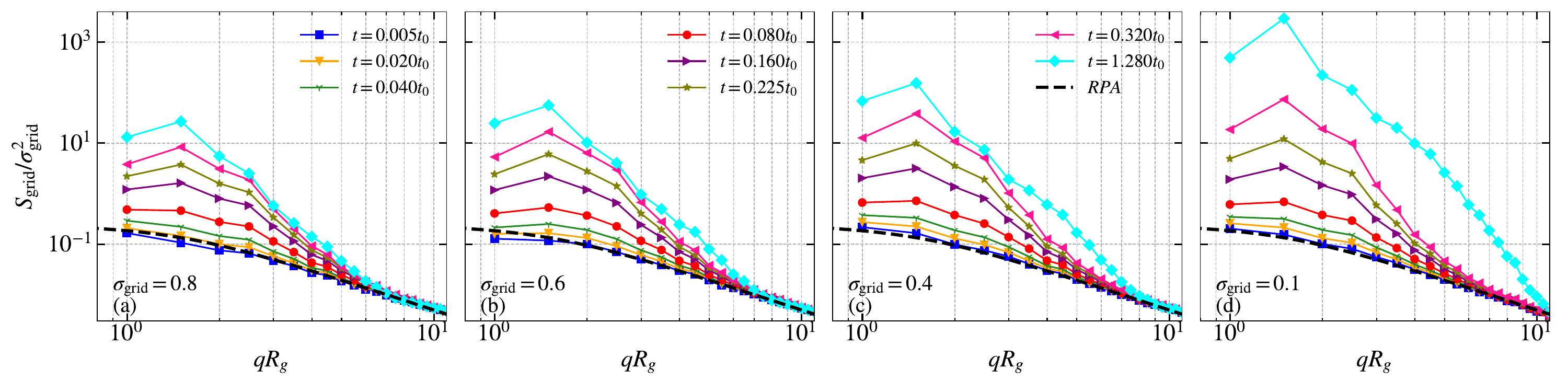} 
   \includegraphics[width=0.9\textwidth]{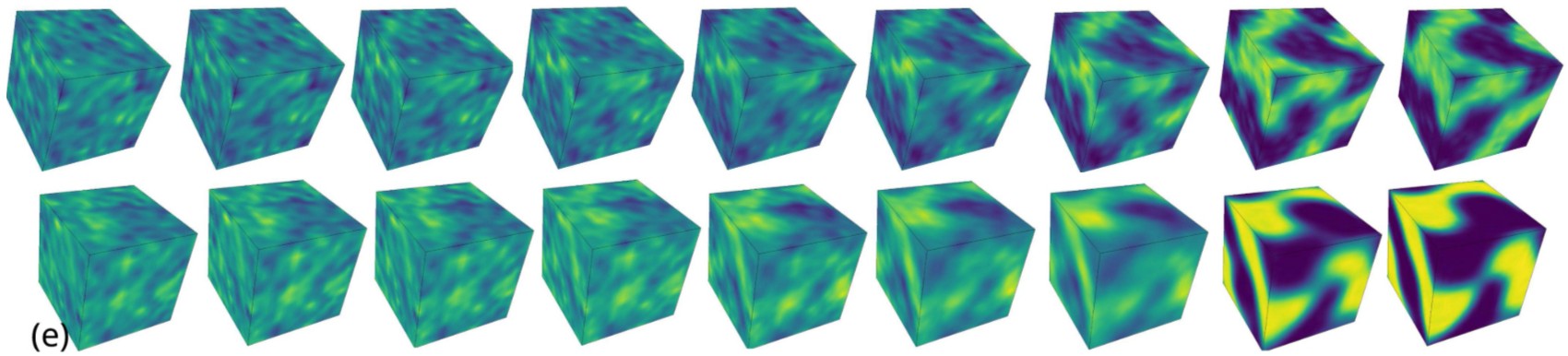}   
   \caption{Dynamics of spinodal decomposition in symmetric polymer
   blends following a quench from $\chi N = 0$ to $\chi N = 28$,
   obtained from DSCFT simulations with noise and mobilities
   according to the reptation model. Panels (a-\rdel{c}\rev{d})
   show the reduced structure factor $\Sgrid(q)/\ssgrid$ 
   at times ranging from $t=0.005 \tO$ to $t=1.28 \tO$ (see legend)
   for noise levels (a) $\sgrid = 0.8$, (b) $\sgrid = 0.6$,\rdel{and} 
   (c) $\sgrid = 0.4$\rev{, and (d) $\sgrid = 0.1$}.  Panel
   (\rdel{d}\rev{e}) shows snapshots for noise levels 
   $\sgrid = 0.8$ (top) and $\sgrid = $\rdel{0.4}$\rev{0.1}$ (bottom) at times 
   $/\tO = 0.005,0.02, 0.04, 0.08, 0.16, 0.22, 0.32, 0.64, 1.28$ 
   (from left to right). 
    \label{fig:Sq-blend-noise}
    }
\end{figure*}

\subsection{Fully disordered configurations}

We begin with the case $\chi N = 0$. In this limit, the MD results for
the scaled structure factor $\frac{1}{N}S(q)$ are in excellent
agreement with the RPA prediction, as shown in \cref{Fig:Sq-blend} and
\cref{Fig:Rq-blend}. The corresponding DSCFT calculations were
initiated with random initial configurations, and we analyzed the 
configurations after reaching steady state for different noise
amplitudes $\sgrid$.  At $\sgrid = 0$, the initial density
fluctuations decay with time and the DSCFT equations eventually yield
completely uniform density fields.  Introducing stochastic noise
generates correlated density fluctuations, whose amplitude increases
with increasing $\sgrid$.  \cref{Fig:Sq-noise-chiN0} shows the resulting
reduced grid structure factor, $\Sgrid(q)/\ssgrid$, for a symmetric
homopolymer blend at $\chi N = 0$ and several values of $\sgrid$. The
curves for different $\sgrid$ do not collapse, indicating that the
noise strength affects the structure factor beyond a simple overall
scaling with $\ssgrid$. In the limit $\sgrid \to 0$, the results are
well described by the RPA prediction (\cref{Eq:RPA}), as indicated by
the solid line. For larger noise strengths, however, the curves
deviate from the RPA expression and the structure
factor is better fitted by the heuristic expression $\Sgrid(q)\approx
\frac{g(q) \ssgrid}{4+6 g(q)\ssgrid}$. 
The inset of \cref{Fig:Sq-noise-chiN0} shows how the reduced structure
factor evolves over time towards this functional form for
$\ssgrid=0.8$, starting from a random initial configuration.

We conclude that introducing noise in the DSCFT equations does enable
the generation of DSCFT configurations with realistic equilibrium
correlations as expected, but only if the noise amplitude is not too
high. For larger noise amplitudes, large-wavelength fluctuations are
underrepresented.  The reason is that the stochastic term in DSCFT
does not capture all fluctuations present in the underlying, formally
exact fluctuating field theory. This can be seen from the derivation
of the SCFT free-energy functional, which involves a partial
saddle-point approximation~\cite{schmid1998}. At steady state
(equilibrium), the stochastic DSCFT equations generate the
distribution
\begin{equation}
\mathscr{P}[\phi_\alpha] \propto 
\exp \left[ - F[\phi_\alpha]/(\ssgrid v_g) \right],
\label{eq:p_rho}
\end{equation}
where $F[\phi_\alpha]$ is the reduced free energy functional,
\cref{eq:energy-scf}. For small $\sgrid$, a Gaussian approximation
can be applied, yielding $\Sgrid(\qq) = \ssgrid \Gamma_2^{-1}$ in the
disordered state, which agrees with the prediction of the random phase
approximation. For larger values of $\sgrid$, however, where higher-order
correlations become important, the distribution (\ref{eq:p_rho})
deviates from the true equilibrium distribution. In the
limiting case $\ssgrid \to \infty$, \cref{eq:p_rho} predicts a 
nearly uniform distribution which not only ignores all connectivity 
effects, but also any further constraints on physically accessible 
density fields such as positivity.

\begin{figure*}[t]
    \centering
    \includegraphics[width=0.9\textwidth]{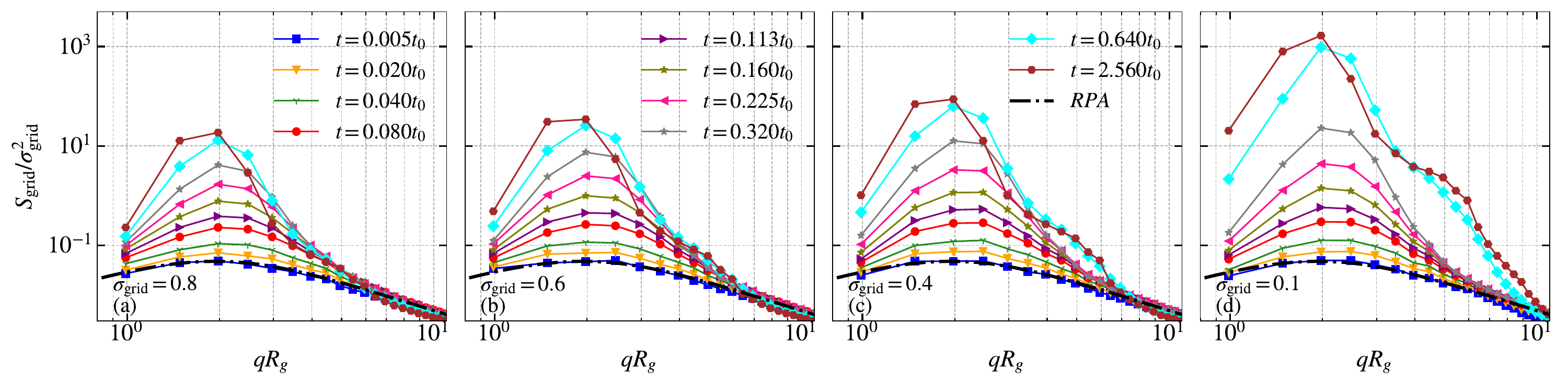} 
   \includegraphics[width=0.9\textwidth]{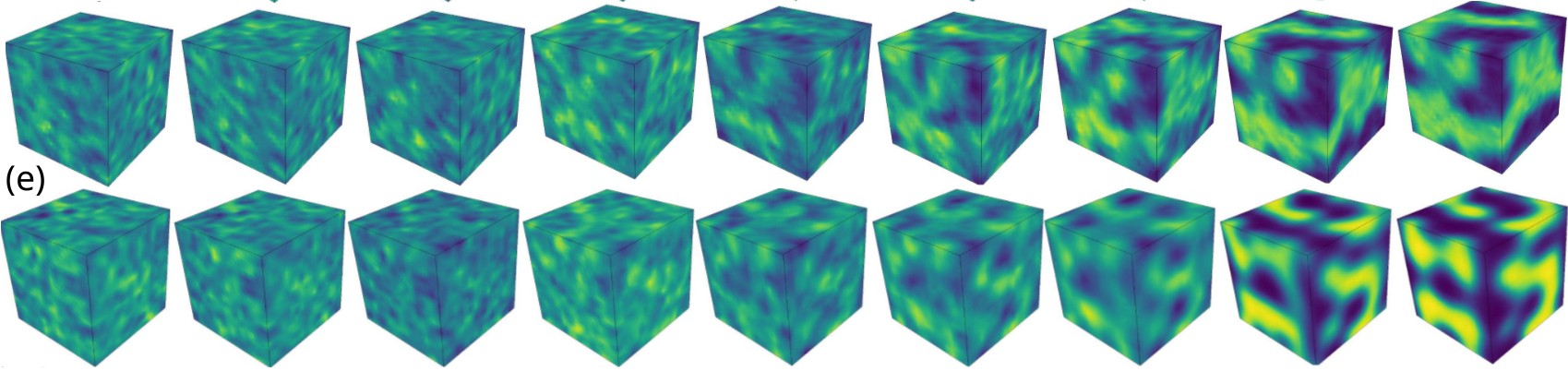}  
    \caption{Dynamics of microphase separation in diblock copolymer
    melts following a quench from $\chi N = 0$ to $\chi N = 20$,
    obtained from DSCFT simulations with noise and mobilities
    according to the reptation model. Panels (a-\rdel{c}\rev{d}) show the
    reduced structure factor $\Sgrid(q)/\ssgrid$ during spinodal
    decomposition at times ranging from $t=0.005 \tO$ to $t=2.56 \tO$
    as indicated and for noise levels (a)$\sgrid = 0.8$, (b) $\sgrid =
    0.6$, \rdel{and} (c) $\sgrid = 0.4$\rev{, and (d) $\sgrid = 0.1$}.
    Panel (\rdel{d}\rev{e}) shows snapshots for noise
    levels $\sgrid = 0.8$ (top) and $\sgrid =$ \rdel{0.4}$\rev{0.1}$ (bottom) at
    times $t/\tO=0.02,0.04, 0.08, 0.113, 0.16, 0.225, 0.32, 0.64,
    2.56$ (from left to right).
    \label{fig:Sq-diblock-noise}
    }
\end{figure*}

\begin{figure}[h!tb]
    \centering
        \includegraphics[width=0.35\textwidth]{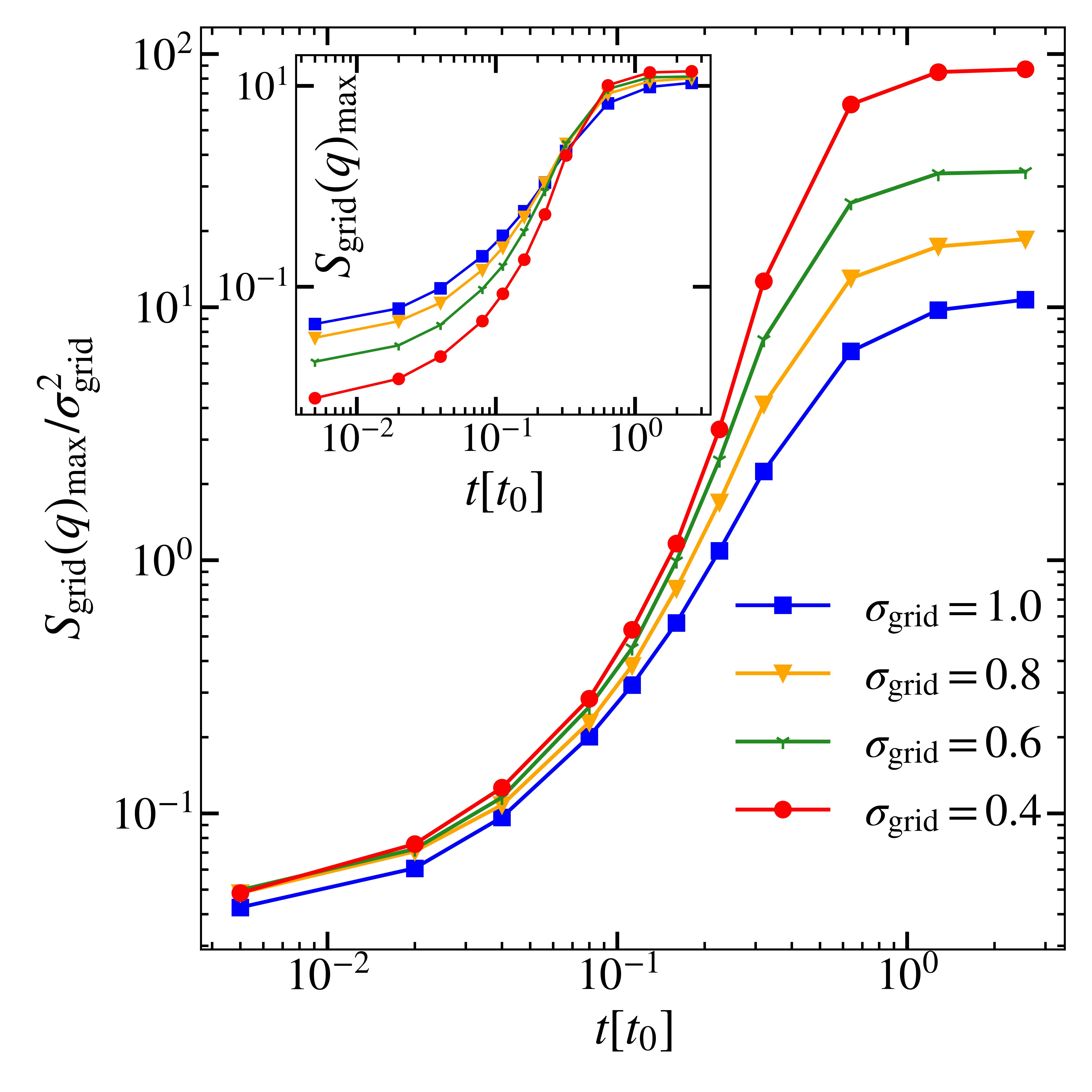}
    \hfill
    \caption{Maximum value of reduced structure factor vs. time
    for microphase separating diblock copolymer melts 
    ($\chi N = 20$) at different noise levels $\sgrid$ as indicated. 
    Inset shows same data for the bare grid structure factor
    $\Sgrid(q)$.
    }
    \label{Fig:Sq-diblock-max-noise}
\end{figure}

Unfortunately, this analysis also implies that systems with
fluctuation levels characteristic of our simulated systems, $\sgrid
\sim 5$ and $\sgrid \sim 2.5$ (for voxel volume $v_g = (0.3 \Rg)^3$
and chain lengths $N=100$  and $N=400$, respectively), cannot be
reliably described within stochastic DSCFT. In the following
discussion, we therefore restrict ourselves to small noise amplitudes,
corresponding to very long polymer chains.
Therefore, unless stated otherwise, we use the mobility
function derived from the reptation model, Eqs.\ (\ref{Eq:esc-rep})
and (\ref{Eq:delta-rep}) in the DSCFT calculations presented below.

\subsection{Spinodal decomposition}

To analyze the effects of noise on DSCFT simulations
(\cref{Eq:DDFT-phi}), we first generated initial configurations at
$\chi N = 0$ for different fluctuation levels. These configurations
were then used as starting points for stochastic DSCFT simulations of
spinodal decomposition in symmetric homopolymer blends and diblock
copolymer melts, following a quench into the two-phase region and
microphase-separated regime, respectively.  Figs.\
\ref{fig:Sq-blend-noise} and (\ref{fig:Sq-diblock-noise}) show the
corresponding reduced structure factor, $\Sgrid(q)/\ssgrid$ 
after a sudden increase of the interaction parameter from
$\chi N = 0$ to $\chi N = 28(20)$ at time $t = 0$ for three
different noise levels, $\sgrid = 0.8$, $0.6$, and $0.4$.

\begin{figure*}[t]
    \centering
    \includegraphics[width=0.9\textwidth]{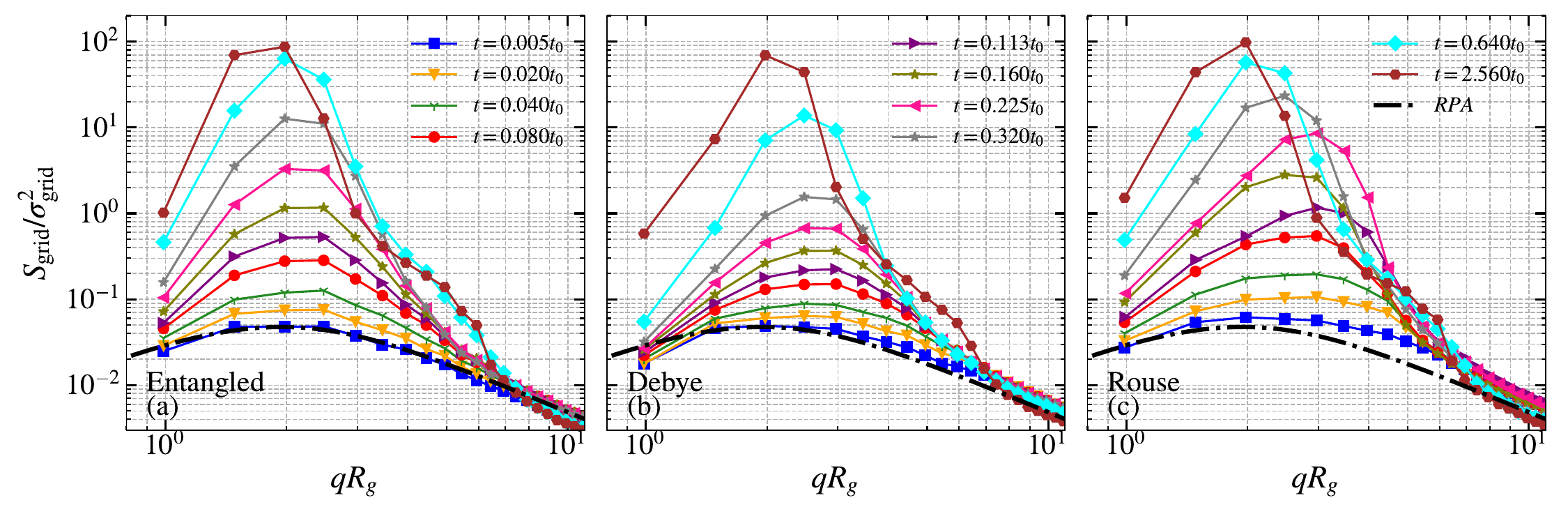}
    \caption{Evolution of reduced structure factor
    $\Sgrid(q)/\ssgrid$ during microphase separation after a
    quench from $\chi N = 0$ to $\chi N = 20$ obtained from
    DSCFT simulations with different mobility functions:
    (a) Reptation model, (b) Debye model, (c) Rouse model. 
    The noise level is $\sgrid=0.4$.
     }
    \label{Fig:Sq_noise_diff_mobilities}
\end{figure*}

\begin{figure*}[!htb]
    \centering
     \includegraphics[width=0.9\textwidth]{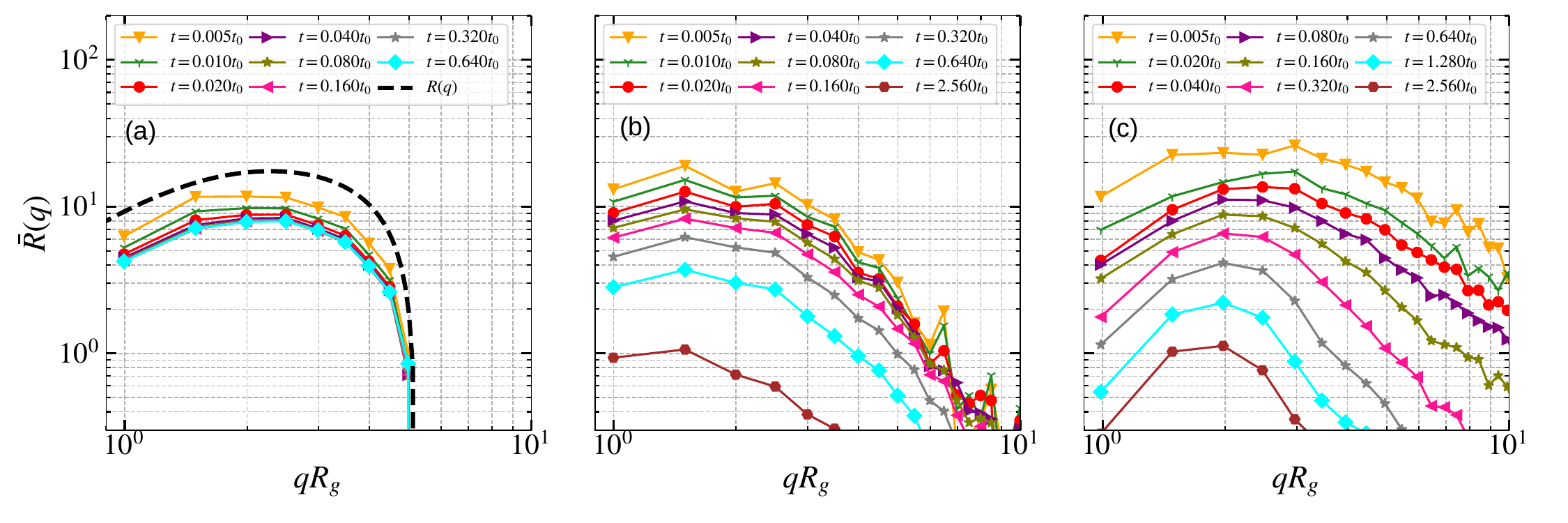}

     \caption{Effect of noise on the average growth rate $\bar{R}(q)$
     during spinodal decomposition
     from DSCFT simulations with mobilities according to
     the reptation model.  
     Panel (a) presents $\bar{R}(q)$ for a
     homopolymer blend with initial configurations prepared using a
     noise strength of $\sigma = 0.01$, while the subsequent evolution 
     ($t > 0$) proceeds without noise  following a quench to $\chi N = 28$.  
     Panels (b) and (c) present $\bar{R}(q)$ for a blend and a copolymer melt
     following a quench to $\chi N = 28$ and $20$, respectively, with
     a noise strength of $\sigma = 1.0$ applied both in generating the
     initial configurations and throughout the simulation of the
     spinodal decomposition.  
     }
    \label{Fig:Rq-noise} 
\end{figure*}

Adding noise affects the evolution of the structure factor in several
ways. First, the unphysical rapid decrease of the structure factor at
high $q$, which can be seen in \cref{Fig:Sq-blend}, is no longer
observed. As anticipated, noise generates and stabilizes the
small-scale density fluctuations that characterize the large-$q$
behavior of the reduced structure factor. Second, the time evolution
of the reduced structure factor at intermediate and late times depends
on the noise level. For small $\sgrid$, a shoulder develops in
the high $q$ regime,indicating the formation of persistent small-scale
structures that do not disappear over time.  This is observed both in
blends (\cref{fig:Sq-blend-noise}c\rev{,d}) and copolymer melts
(\cref{fig:Sq-diblock-noise}c\rev{,d}). At higher noise levels (Figs.\
\ref{fig:Sq-blend-noise}a) and \ref{fig:Sq-diblock-noise}a), the
shoulder disappears. We attribute it to local long-lived defects,
which anneal more easily if activated processes are facilitated by
stronger fluctuations. Indeed, the snapshots in Figs.\
\ref{fig:Sq-blend-noise}\rdel{d}\rev{e}) and
\ref{fig:Sq-diblock-noise}\rdel{d}\rev{e}) show that
some domains are still connected by narrow necks if the noise level is
low (bottom row, $\sgrid = $\rdel{0.4}$\rev{0.1}$), whereas the necks dissolve at higher
noise level (top row, $\sgrid = 0.8$).  

At late times, when the compositions inside the $A$ and $B$
(micro)domains have reached saturation, $\Sgrid(q)$ becomes
independent of $\ssgrid$. Consequently, the reduced structure factor
$\Sgrid(q)/\ssgrid$ becomes larger in systems with lower noise levels,
both in polymer blends and in block copolymer melts.
\cref{Fig:Sq-diblock-max-noise} presents the peak values of the reduced
structure factor in copolymer melts (cf.\ \cref{fig:Sq-diblock-noise})
as a function of time for different noise strengths.  At early times,
the curves are close to each other, but at late times, they deviate
from each other, and saturate at different values. If one instead
plots $\Sgrid(q)$ without normalization (inset of
\cref{Fig:Sq-diblock-max-noise}), the curves converge at late time,
demonstrating that the structural characteristics of systems with
different fluctuation levels become increasingly similar.  

\cref{Fig:Sq_noise_diff_mobilities} illustrates the influence of the
choice of mobility matrix on the evolution of the reduced structure
factor $\Sgrid(q)/\ssgrid$ of diblock copolymer melts during spinodal
decomposition. The system is quenched from $\chi N = 0$ to $\chi N =
20.0$, and the noise strength is $\sgrid = 0.4$. \rev{Such a low noise 
level corresponds to a highly entangled polymer system, where the 
reptation model is most appropriate. Nevertheless, we also discuss
the Rouse model here to highlight the differences between models.}
The structure factor grows fastest if the system evolves with the
Rouse mobility matrix. This can be attributed to the fact that the
function $\lhat_\text{AA}(q)$ is larger in the Rouse model than in the
reptation model and the Debye scheme across all wave vectors $q$.
Furthermore $\lhat_\text{AA}(q)$ is larger in the reptation model than
in the Debye scheme at low $q$, which generally leads to a faster
growth of the structure factor in the reptation model.

\begin{figure*}[t]
    \centering
    \includegraphics[width=0.9\textwidth]{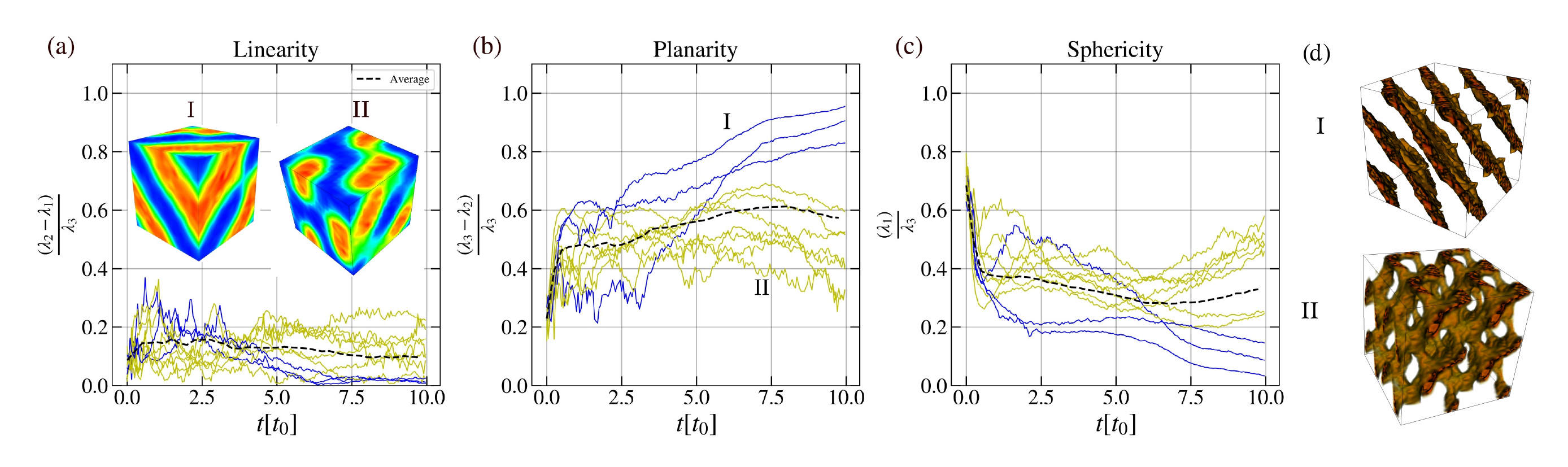} 
    \caption{Eigenvalue analysis of the structure tensor for 10 different
    trajectories of spinodal decomposition in copolymer melts
    at noise level $\sgrid = 0.4$. Panels (a),(b), and (c) show
    linearity, planarity, and sphericity, respectively. The dashed
    lines show the average of the samples.  Inset in a): Images I and II
    correspond to the last configurations of the curves labelled I and
    II in (b). Panel (d) shows snapshots of the same system
    with $g_s \approx 0$ (I) and $g_s \approx 0.6$ (II),
    periodically replicated in all directions and showing only the 
    A-rich block.
    }
    \label{fig:structure-tensor}
\end{figure*}

Next, we discuss the effect of noise on the average growth rate of the
structure factor, $\bar{R}(q)$, defined in \cref{Eq:Rt}.
\cref{Fig:Rq-noise} shows $\bar{R}(q)$ at different times during
spinodal decomposition in polymer blends and copolymers, following a
quench from $\chi N = 0$ to $\chi N = 28$.  \cref{Fig:Rq-noise}a)
shows $\bar{R}(q)$ for a blend whose initial configurations (for
20 independent runs )
were prepared with a noise strength of $\sigma = 0.01$, while the
subsequent evolution ($t > 0$) proceeds deterministically according to
\cref{Eq:DDFT-phi-deterministic}.  In this case, the growth rate
remains nearly constant over a range of times and is well described by
the linearized theory of spinodal decomposition.  \cref{Fig:Rq-noise}b
and \cref{Fig:Rq-noise}c) display $\bar{R}(q)$ for a blend and a
copolymer melt, respectively, as obtained from stochastic DSCFT
simulations with noise strength of $\sigma = 1.0$ following a quench 
to $\chi N = 28$ and $20$, respectively.  In the absence of
noise, the growth rate $\bar{R}(q)$ equals zero at a critical wave
vector $q_\text{c}$ and becomes negative for $q > q_\text{c}$. This
behavior is not observed in the stochastic simulations with noise.
Furthermore, $\bar{R}(q)$ varies slowly with time even for small 
$t$, hence the structure factor growth rate is never purely
exponential.

In DSCFT simulations, we can access late stages of spinodal
decomposition, where strongly entangled systems may get trapped in
different competing morphologies. To examine this effect, we consider
the structure tensor, an alternative observable which is popular in
the context of image analysis. A structure tensor is a $D$-dimensional
field matrix, which is derived from the gradient of an image or a field.  
In three dimensions, it is defined as a matrix $\mathbf{I}$ with
components $I_{ij} = \left< \nabla_i \psi \, \nabla_j \psi \right>$,
where $\nabla_i$ represents the gradient along the $i$-th direction, 
$\langle \cdot \rangle$ denoting an average over data points, and 
$i,j = x,y,z$.  The structure tensor provides a quantitative measure of the
average orientation and anisotropy of spatial patterns in the system.
Specifically, the ratios of its eigenvalues $\lambda_1 \leq \lambda_2
\leq \lambda_3$ are used to quantify the linearity ($g_l$), planarity
($g_p$), and sphericity ($g_s$) of spatial patterns 
as follows~\cite{westin2002processing}: 

\begin{equation} g_l =
\frac{\lambda_2-\lambda_1}{\lambda_3}, \ \  g_p =
\frac{\lambda_3-\lambda_2}{\lambda_3}, \ \ g_s =
\frac{\lambda_1}{\lambda_3} 
\end{equation}

\cref{fig:structure-tensor} shows the ratios of the
eigenvalues of the structure tensor for ten independent runs of a
copolymer melt with noise strength $\sgrid = 0.4$.  For the lamellar
phase, one expects $g_p = 1.0$ and $g_l = g_s = 0$. Only three out of
ten runs approach this state, which corresponds to the free energy
minimum. The other runs remain trapped in metastable configurations
corresponding to a local free energy minimum. A closer inspection
reveals two dominating structures (I,II in insets of
\cref{fig:structure-tensor}a and in \cref{fig:structure-tensor}
d)).  The first, I, is lamellar ($g_s= 0$) with diagonally oriented
lamellae that are connected across periodic boundaries. The second, II
($g_s \sim 0.6$), is a bicontinuous network structure, reminiscent of
a gyroid phase.  Although such structures are metastable, they
are observed for all noise levels considered here. Once they have
fully developed, the systems no longer escapes to the true equilibrium
state.  We should note that, in the present case, the network
structures are stabilized by the periodic boundary conditions and the
finite size of the system, and might be less abundant in larger
system.  Nevertheless, these results demonstrate that systems evolved
by stochastic DSCFT do not necessarily relax to the true equilibrium
state, but can remain kinetically trapped in metastable
configurations. At the same time, they highlight the ability of DSCFT
to investigate process-dependent non-equilibrium structure formation.

\section{Summary and Conclusions}

We have refined and tested a dynamic self-consistent field theory
(DSCFT) for polymer blends and diblock copolymer melts in which the
nonlocal mobility matrix is constructed from the relaxation dynamics
of single-chain structure factors. This approach extends previous
work\cite{mantha2020bottom,schmid2020dynamic} on unentangled polymers
to the entangled regime.

First, mobility functions were determined from molecular dynamics
simulations of homogeneous melts covering chain lengths from the
unentangled to the moderately entangled regime. For highly entangled
chains, analytical expressions were derived using the reptation
theory.  The resulting mobilities reveal substantial deviations from
the commonly used Debye approximation. In particular, entanglement
effects strongly suppress mobility at intermediate length scales, and
the mobility matrix of diblock copolymers -- both diagonal and
off-diagonal components -- exhibits nontrivial contributions
associated with internal chain dynamics.

Second, the mobility functions were incorporated into DSCFT
calculations of spinodal decomposition in symmetric homopolymer blends
and ordering kinetics in symmetric diblock copolymer melts. To
describe blends, the theory was generalized so that local fluxes
depend on local chain densities. Comparisons with molecular dynamics
simulations showed that DSCFT captures the overall evolution of domain
structures and characteristic length scales. Mobility functions
derived from single-chain dynamic structure factors consistently
outperform Debye-type mobilities and yield improved agreement with
simulation results, especially for diblock copolymers. Nevertheless,
systematic discrepancies remain at very short times, consistent
with literature findings\cite{steffen2025collective} that
memory effects neglected in the present Markovian formulation contribute 
to the initial stages of structure formation.

Third, we investigated the influence of stochastic currents. Thermal
noise provides a practical route for generating equilibrium disordered
initial configurations and eliminates artifacts associated with
mapping particle configurations onto density fields. Noise also
counteracts the unphysical decay of large-wavelength fluctuations 
during spinodal decomposition, which is observed in deterministic DSCFT
calculations, and can facilitate access to alternative kinetic
pathways. However, our analysis also shows that large noise levels
result in improper sampling of the configuration space, especially
with respect to long-wavelength correlations.

Overall, the present work establishes a quantitatively grounded
DSCFT framework that incorporates realistic polymer dynamics from
microscopic simulations and analytical reptation theory. The approach
provides an efficient alternative to particle-based simulations for
studying structure formation in blends of polymers and block
copolymers with high molecular weight over experimentally relevant
length and time scales, including systems with significant chain
entanglement.

The present framework opens several directions for further
development. A natural next step is to extend the theory beyond its
current Markovian formulation by incorporating memory effects in the
mobility kernel, along the lines of the works by M\"uller et
al\cite{wang2019collective,rottler2020kinetic,steffen2025collective},
in order  to analyze the effects of entanglements on the earliest
stages of (micro)phase separation.  Another important avenue is the
refinement and generalization of mobility functions for more complex
polymer architectures and conditions, including dynamically asymmetric
block copolymers, polydisperse systems, branched polymers, or
reversible networks. In these cases, connectivity effects and
topological constraints are expected to produce even stronger nonlocal
and history-dependent contributions to the dynamics, requiring either
improved theoretical descriptions or additional input from
simulations.  From a methodological perspective, coupling DSCFT more
tightly with coarse-grained particle simulations or data-driven
approaches offers a promising route to systematically improve the
mobility descriptions. Machine-learned or adaptive mobility kernels,
trained on targeted molecular simulations, could provide a practical
way to include missing physics while retaining computational
efficiency at the field level.  These developments would further
strengthen DSCFT as a multiscale bridge between microscopic polymer
physics and mesoscale pattern formation, moving toward a more complete
and predictive field-based description of polymer dynamics.

\rev{
\section*
{Supporting Information}
Supporting information file contains
\begin{description}[noitemsep,topsep=0pt]
\item[SI-1] Analytical considerations on the
asymptotic behavior of single-chain mobility functions in the 
long-wavelength limit (see Eq.\ \eqref{eq:gqt_asym}). 
\item[SI-2]  Derivation of analytical expressions for the mobility 
functions of block copolymer chains based on the reptation model 
(see \eqref{Eq:delta-rep}). 
\item[SI-3]  Simulation results (with Fig. S1) for the effective 
coordination number in the long-chain limit.
\item[SI-4]  Comparison of structure factors from MD simulations 
and DSCFT calculations in one panel (with Fig. S2) 
\item[SI-5] Derivation of analytical expressions for the 
average growth rate in RPA approximation. 
(See Eqs.\ (\ref{Eq:Rt-linear-homo},ref{Eq:Rt-linear-copolymer}).
\end{description}
}

\section*{Acknowledgements}

This research was supported by the German Science Foundation (DFG) --
Project number 233630050 --  via CRC/TRR 146 (project C1). The
simulations were carried out on the high performance computing center
MOGON at JGU Mainz.

\section*{Conflict of Interest}
The authors declare no competing interests.

\section*{Author contributions}
{\bf Alireza F. Behbahani:} Conceptualization (supporting);
Methodology (equal); Software (supporting); Investigation (equal);
Formal Analysis (equal); Data Curation (equal); Writing - Original
Draft (equal); Writing - Review and Editing (supporting).
{\bf Jafar Cheraghalizadeh:} Methodology (supporting);
Conceptualization (supporting); Software (lead); Investigation
(equal); Formal Analysis (equal); Data Curation (equal); Writing -
Original Draft (equal); Writing - Review and Editing (supporting).
{\bf Friederike Schmid:} Conceptualization (lead); Methodology
(equal); Resources (lead); Writing - Review and Editing (lead);
Supervision (lead); Project Administration (lead); Funding Acquisition
(lead).

\section*{Data Availability}
The data supporting the findings of this paper are openly available in the Zenodo repository at
\url{https://doi.org/10.5281/zenodo.20626607} and
\url{https://doi.org/10.5281/zenodo.20698614}.  \\
The source code used for the simulations and analysis is publicly available at \url{https://github.com/jafarcheraghalizadeh/SF-DDFT.git}.

\bibliographystyle{apsrev4-2-titles}
\bibliography{refs}

% SUPPLEMENTARY INFORMATION

\clearpage

 \renewcommand{\thepage}{S\arabic{page}}
 \renewcommand{\thesection}{SI $-$ \arabic{section}}
 \renewcommand{\thetable}{S\arabic{table}}
 \renewcommand{\thefigure}{S\arabic{figure}}
 \renewcommand{\theequation}{S\arabic{equation}}

 %\renewcommand*{\citenumfont}[1]{S#1}%
 %\renewcommand*{\bibnumfmt}[1]{(S#1)}

%\title{Supporting Information for \\
%Dynamics of Structure Formation in Block Copolymer Melts and Polymer Blends: Dynamical Self-Consistent Field Theory and Molecular Dynamics Simulations}

\section*{Supporting Information}

\section{Asymptotic behavior of single-chain mobility functions 
in the long wavelength limit}

We consider general heteropolymers containing $N$ monomers with $\nu$
different segment types. We denote by $S_\alpha$ the set of monomers of
type $\alpha$, and by $f_\alpha$ the fraction of $\alpha$-monomers in
the entire molecule. Note that the molecules do not have to form linear
chains; we do not specify how the $N$ monomers are connected to each
other. Following the main text, our task is to calculate the rescaled
single-chain mobility

 \begin{equation}
 \label{si-eq:mobility}
 \llhat(\qq)
 = \frac{1}{N^2} \: \bg(\qq,0) \: \GG^{-1}(\qq) \: \bg(\qq,0),
 \end{equation}
 where
 \begin{equation}
 g_{\alpha \beta}(\qq,t) = \frac{1}{N} 
  \left\langle \subsum \: \ue^{\ui \qq \cdot (\rr_n(t) - \rr_m(0))} 
  \right\rangle
 \end{equation}
 is the tensorial single-chain structure factor, and
 \begin{equation}
 \GG(\qq) = \frac{q^2}{N} \int_0^\infty \ud t \: 
 \bg(\qq,t)
 \end{equation}
 its rescaled integral over time. For future reference, we also
 introduce the total single-chain structure factor and its integral
 \begin{equation}
 g(\qq,t) = \sum_{\alpha \beta} g_{\alpha \beta}(\qq,t),
 \quad
 G(\qq) = \frac{q^2}{N} \int_0^\infty \ud t \: g(\qq,t).
 \end{equation}
 
 The single chain structure factor has one single dominant slow
 relaxation mode at small $\qq$ vectors, which corresponds to the
 diffusion of the whole chain.  As a result, the matrix $\GG(\qq)$ is
 singular at $\qq = 0$, and nearly singular at $\qq \to 0$, which
 may cause numerical problems. To deal with them, we must isolate 
 the singularity.

 To this end, we inspect the single chain structure factor more closely.
 First, we perform a cumulant expansion, and get
 \begin{equation}
   g_{\alpha \beta}(\qq,t)  \approx \frac{1}{N} \subsum \:
   \ue^{ - \frac{q^2}{6} \langle (\rr_n(t) - \rr_m(0))^2 \rangle}
 \end{equation}
 This expression is exact for Rouse chains, and more generally valid
 up to order $(q \Rg)^2$.  Next, we split the monomer position vectors
 into $\rr_n = \RR_c + \uu_n$ with the center of mass vector $\RR_c(t)
 = \frac{1}{N} \sum_n \rr_n(t)$ and the relative coordinate vector
 $\uu_n(t) = \rr_n(t) - \RR_c(t)$.  At time $t=0$, the contribution
 of $\RR_C$ drops out and we get
 \begin{eqnarray}
 \nonumber
\lefteqn{ g_{\alpha \beta}(\qq,0) 
 = \frac{1}{N} \subsum 
   \ue^{ - \frac{q^2}{6} \langle (\uu_n - \uu_m)^2 \rangle}} \quad &&
 \\ & = &
 %\\ & \stackrel{q \Rg \ll 1 }{\approx} &
   \frac{1}{N} \subsum 
   \big(1 - \frac{q^2}{6} \langle (\uu_n - \uu_m)^2 \rangle \big)
  + {\cal O}(q^4)
 \nonumber
 \\ & = &
 %\\ & \stackrel{q\Rg \ll 1}{\approx} &
   N I_\alpha(\qq) I_\beta(\qq)
   + \frac{q^2}{3} \frac{1}{N} 
     \subsum \!\! \langle \uu_n \cdot \uu_m \rangle
  + {\cal O}(q^4) 
 \label{si-eq:gq0}
 \end{eqnarray}
 with
 \begin{equation}
 I_\alpha(\qq) =  \frac{1}{N}
 \big\langle \sum_{n \in S_\alpha} \:
 \ue^{\ui \qq \cdot \uu_n} \big \rangle
 =
 f_\alpha \: 
   \ue^{- \frac{q^2}{6} \langle \uu^2 \rangle_\alpha}
   + {\cal O}(q^4)
\label{si-eq:Ialpha}
 \end{equation}
 and $\langle \cdot \rangle_\alpha$ denotes the average
 for monomers of type $\alpha$.

 At late times, the center of mass coordinate and the 
 relative coordinates decorrelate, and we get
\begin{eqnarray}
 \lefteqn{g_{\alpha \beta}(\qq,t) 
 \!\! \stackrel{t \gg \Rg^2/D}{\approx}\!\!\!   
   \ue^{- \frac{q^2}{6} \langle (\RR_c(t)-\RR_c(0))^2 \rangle}}
 \quad
 \nonumber
 \\ && \hspace*{20mm} \times
   \frac{1}{N} \subsum
   \ue^{ - \frac{q^2}{6} \langle (\uu_n(t) - \uu_m(0))^2 \rangle}
 \nonumber
 \\ & =&
   \ue^{- D  q^2 t} \:
   \Big( N \: I_\alpha(\qq) \: I_\beta(\qq)
  \nonumber
  \\ && \quad
   + \: \frac{q^2}{3} \: \frac{1}{N} \subsum
     \langle \uu_n(t) \cdot \uu_m \rangle 
   \Big) + {\cal O}(q^4).
     \label{si-eq:gqt}
 \end{eqnarray}
 Comparing Eqs.\ (\ref{si-eq:gq0}) and (\ref{si-eq:gqt}) with each
 other, we find that both at $t=0$ and in the limit
 $t \gg R_g^2/D$, the function $\bg(\qq,t)$ can be
 approximated by
 \begin{equation}
   g_{\alpha \beta}(\qq,t)  \approx
   N \: \ue^{- D q^2 t} \: 
   \left( I_\alpha(\qq) I_\beta(\qq)
     + \frac{q^2}{3} c_{\alpha \beta}(t) \right)
  + {\cal O}(q^4)
  \end{equation}
 with 
 \begin{equation}
  c_{\alpha \beta}(t) = 
   \frac{1}{N^2}
   \subsum \langle \uu_n(t) \cdot \uu_m(0) \rangle.
 \label{si-eq:c}
 \end{equation}
 The correlation functions $c_{\alpha \beta}(t)$ satisfy
 \begin{equation}
 \sum_{\alpha} c_{\alpha \beta}(t) = 
  \sum_{\beta} c_{\alpha \beta}(t) = 0,
 \label{si-eq:sumc}
 \end{equation}
 since $\sum_n \uu_n(t) = 0$ by definition, and vanish at late 
 times, since $\langle \uu_n(t) \cdot \uu_m(0) \rangle 
 \stackrel{t \to \infty}{\longrightarrow}
 \langle \uu_n \rangle \: \langle \uu_m \rangle = 0$.
 The integral
 \begin{equation}
 C_{\alpha \beta}(q) := 
 \int_0^\infty \ud t \: \ue^{-D q^2 t} c_{\alpha \beta}(t) 
 \stackrel{q \to 0}{\approx} 
 \int_0^\infty \ud t \: c_{\alpha \beta}(t)
 \end{equation}
 is a finite matrix at $q \to 0$.

 For small $(q\Rg)$, the integral
 $\int_0^\infty \ud t \: \bg(\qq,t)$ is governed by the late
 time behavior, and we get
 \begin{displaymath}
 \frac{1}{N} \int_0^\infty \ud t \: g_{\alpha \beta}(\qq,t)
 \approx \frac{1}{D q^2} I_\alpha(\qq) I_\beta(\qq)
  + \frac{q^2}{3} \: C_{\alpha \beta}(q)
  + {\cal O}(q^4).
 \end{displaymath}

 The quantities entering Eq.\ (\ref{si-eq:mobility}) are thus given by
 \begin{eqnarray}
 \label{si-eq:gg}
 \frac{1}{N} \bg(\qq,0) & =& 
  \II(q) \II^T(q)  + \frac{q^2}{3} \cc(0) + {\cal O}(q^4)
 \\
 \GG(\qq) &=& \frac{1}{D} \: \II(q) \II^T(q)
 + \frac{q^4}{3} \: \CC(q)
 + {\cal O}(q^6).
 \label{si-eq:GG}
 \end{eqnarray}
 At $q = 0$, we thus have $\GG(\qq) \propto \II(q) \II^T(q)$,
 which is a projection operator and hence a singular matrix
 that cannot be inverted. Nevertheless, Eq.\ (\ref{si-eq:mobility})
 can be evaluated at finite $q$ and the limit $q \to 0$ is finite.

 To analyze this limit, we switch to a basis in ``component
 space'' where the relations $\sum_{\alpha} c_{\alpha \beta}(t) = 
 \sum_{\beta} c_{\alpha \beta}(t) = 0$ can be exploited: 
 The
 first basis vector is the $\nu$-dimensional unit vector $\ee_1 =
 (1,..., 1)/\sqrt{\nu}$. The other basis vectors $\ee_2, ...,\ee_\nu$
 are orthogonal to $\ee_1$.  Then we have
 \begin{equation}
 \label{si-eq:ctot}
 \ee_1^T \: \cc(t) \: \ee_j = \ee_j^T \: \cc(t) \: \ee_1 = 0,
 \end{equation}
 for all $t$ and $j$,
 since $\sum_\alpha c_{\alpha \beta}(t) = \sum_\beta c_{\alpha
 \beta}(t) = 0$ as discussed above, and
 \begin{equation}
 \label{si-eq:gtot}
 \ee_1^T \: \bg(\qq,t) \: \ee_1
 = \frac{1}{\nu} \sum_{\alpha \beta} g_{\alpha \beta}(\qq,t) 
 = \frac{1}{\nu} \: g(\qq,t).
 \end{equation}
 We rewrite Eqs.\ (\ref{si-eq:gg}) and
 (\ref{si-eq:GG}) in the new basis for the matrix elements
 $\bar{g}_{ij}(\qq,0) = \ee_i^T \bg(\qq,0) \ee_j$ and
 $\bar{G}_{ij}(\qq) = \ee_i^T \GG(\qq) \ee_j$:
 \begin{eqnarray}
 \label{si-eq:gg2}
 \frac{1}{N} \bar{g}_{ij}(\qq,0) & =& 
  \bar{I}_i(q) \bar{I}_j(q)  
    + \frac{q^2}{3} \bar{c}_{ij} + {\cal O}(q^4)
 \\
 \bar{G}_{ij}(\qq) &=& \frac{1}{D} \: 
  \bar{I}_i(q) \bar{I}_j(q)  
 + \frac{q^4}{3} \bar{C}_{ij}(q)
 + {\cal O}(q^6),
 \label{si-eq:GG2}
 \end{eqnarray}
 with $\bar{I}_i = \ee_i^T  \II, \;
 \bar{c}_{ij} = \ee_i^T \cc(0) \: \ee_j$, and
 $\bar{C}_{ij}(q) = \ee_i^T \CC(q) \: \ee_j$.
 Specifically, Eq.\ (\ref{si-eq:ctot}) results in
 \begin{displaymath}
 \bar{c}_{1j} = \bar{c}_{j1} = \bar{C}_{1j}(q) = \bar{C}_{j1}(q)= 0,
 \end{displaymath}
 which allows us to rewrite
 $\bar{\cc}$,  $\bar{\CC}(q)$ as
 \begin{displaymath}
 \bar{\cc} =: \xemat{0}{0}{0}{\mat{c}}, \quad
 \bar{\CC}(q) =: \xemat{0}{0}{0}{\mat{C}(q)}, 
 \end{displaymath}
 where $\mat{c}$ and $\mat{C}(q)$ are ($\nu-1$) dimensional
 matrices in the subspace spanned by $(\ee_2, ... \ee_{\nu})$.
 Furthermore, rewriting $\bar{\II}$ as $(\bar{I}_1, \vvec{I})$
 with $\vvec{I} = (\bar{I}_2, ... \bar{I}_\nu)$, and using
 Eq.\ (\ref{si-eq:gtot}), we can relate $\bar{I}_1(q)$ to
 $g(\qq,0)$ and $G(\qq)$ up to order $q^4$:
 \begin{eqnarray*}
 \frac{1}{N} \bar{g}_{11}(\qq,0) 
   &=& \frac{\nu}{N} \: g(\qq,0)
   = \bar{I}_1(q)^2  + {\cal O}(q^4)
\\
 \bar{G}_{11}(\qq) 
   &=& \nu \: G(\qq)
   = \frac{1}{D} \bar{I}_1(q)^2   + {\cal O}(q^6).
 \end{eqnarray*}
 The matrix equations for $\bar{\bg}(\qq,0)$ and $\bar{\GG}(\qq)$
 turn into
 \begin{eqnarray*}
 \frac{1}{N} \bar{\bg}(\qq,0) & =& 
  \xemat{\bar{I}_1^2(q)}{\bar{I}_1(q) \: \vvec{I}^T(q)}
    {\bar{I}_1(q) \vvec{I}(q)} {\vvec{I}(q) \: \vvec{I}^T(q)}
\\ && \quad
     + \frac{q^2}{3} 
     \xemat{0}{0}{0}{\mat{c}}
    + {\cal O}(q^4)
 \\ \nonumber
 \bar{\GG}(\qq) &=& \frac{1}{D} \: 
  \xemat{\bar{I}_1^2(q)}{\bar{I}_1(q) \:\vvec{I}^T(q)}
    {\bar{I}_1(q) \:\vvec{I}(q)} {\vvec{I}(q) \: \vvec{I}^T(q)}
\\ && \quad
 + \frac{q^4}{3} 
   \xemat{0}{0}{0}{\mat{C}(q)}
 + {\cal O}(q^6).
 \end{eqnarray*}
 Written in this way, we can explicitly invert $\bar{\GG}$,
 %\begin{displaymath}
 %\bar{\GG}^{-1}(\qq) \approx \frac{3}{q^4 \:\bar{I}_1^2(q)} \:
 %\xemat{\frac{q^4}{3} \: D + 
 %        \vvec{I}^T(q) \mat{C}^{-1}(q) \:\vvec{I}(q)}
 %{- \bar{I}_1(q) \: \vvec{I}^T(q) \mat{C}^{-1}(q)}
 %{- \bar{I}_1(q) \: \mat{C}^{-1}(q) \: \vvec{I}(q)}
 %{ \bar{I}_1^2 (q)\: \mat{C}^{-1}(q)} \: (1 + {\cal O}(q^2))
 %\end{displaymath}
 and after some algebra, obtain the following simple
 expression for the mobility matrix in the new basis:
 \begin{eqnarray*}
 \bar{\Lambda}(\qq) 
 &=& \frac{1}{N^2} \: \bar{\bg}(\qq,0) \: 
  \bar{\GG}^{-1}(\qq) \: \bar{\bg}(\qq,0)
  \nonumber \\
 &=& 
 D \: \bar{\II}(q) \bar{\II}^T(q)
 + \frac{1}{3} 
 \xemat{0}{0}{0}{\mat{c} \: \mat{C}^{-1}(q) \: \mat{c}}
 + {\cal O}(q^2).
 \end{eqnarray*}
 In the original basis, the above equation reads
 \begin{equation}
 \Lambda(\qq) 
 = D \II(q) \II^T(q)
 + \frac{1}{3} \cc(0) \: \CC^+(q) \: \cc(0)
 + {\cal O}(q^2)
 \end{equation}
 where $\CC^+(q)$ is the pseudoinverse of $\CC(q)$.
 Using \eqref{si-eq:sumc} and $\sum_{\alpha} f_\alpha = 1$, we
 recover $\sum_{\alpha \beta} \Lambda_{\alpha \beta}(0) = D$
 as expected.

\section{Mobility functions of block copolymer chains based on the reptation model}

For chains undergoing pure reptation, Doi and
Edwards\cite{doi1988theory} derive the following expression for the
correlation function  $\phi(s,s';t) = \langle \exp\{\ui
\qq \cdot [\rr(s,t) - \rr(s',0)] \} \rangle$, where $s$
denotes the contour distance from a chain end:
\begin{equation}
\begin{aligned}
\phi(s,s';t)
=& \sum_{p=1}^{\infty}
\Biggl[
\frac{2\mu}{\mu^2 + \alpha_p^2 + \mu}
\cos\!\left[\frac{2\alpha_p}{L}\left(s-\frac{L}{2}\right)\right]
\\[4pt] & \qquad \times
\cos\!\left[\frac{2\alpha_p}{L}\left(s'-\frac{L}{2}\right)\right]
\exp\!\left(-\frac{4 \Dc  \alpha_p^2}{L^2}\, t\right)
\\[4pt] &
+
\frac{2\mu}{\mu^2 + \beta_p^2 + \mu}
\sin\!\left[\frac{2\beta_p}{L}\left(s-\frac{L}{2}\right)\right]
\\[4pt] & \qquad \times
\sin\!\left[\frac{2\beta_p}{L}\left(s'-\frac{L}{2}\right)\right]
\exp\!\left(-\frac{4 \Dc  \beta_p^2}{L^2}\, t\right)
\Biggr].
\end{aligned}
\end{equation}
Here $L = Z a$ is the contour length of the primitive chain (or contour
length of the tube),  $Z = N/N_\mathrm{e}$ is the number of steps
of the tube and $a$ is the step length of the tube.  The quantity
$\Dc = k_\mathrm{B}T/(N\zeta)$ is the curvilinear diffusion
coefficient of the chain along the tube.  The parameter $\mu = q^2
R_\text{ee}^/12$ and $\alpha_p$ and $\beta_p$ are the positive solutions
of the equations $\alpha_p \tan \alpha_p = \mu$ and $\beta_p \cot
\beta_p = -\mu$, respectively.  

Based on the above correlation function, partial single-chain dynamic structure factors for a symmetric diblock copolymer melt can be calculated using:
\begin{equation}
    \begin{aligned}
\lefteqn{g_{\mathrm{AA}}(q,t) = 
\frac{N}{L^2} \int_0^{L/2} \! \int_0^{L/2} \phi(s,s';t) \, 
   \mathrm{d}s \, \mathrm{d}s' } \quad &
\\[4pt] & 
= N \sum_{p=1}^{\infty} \Biggl[
\frac{2 \mu}{\mu^2 + \alpha_p^2 + \mu} \, 
\frac{\sin^2 \alpha_p}{4 \alpha_p^2} \, 
\exp\Biggl(- \frac{4 D_\mathrm{c} \alpha_p^2}{L^2} t \Biggr) 
\\[4pt] & \qquad 
+ \frac{2 \mu}{\mu^2 + \beta_p^2 + \mu} \, 
\frac{(1 - \cos \beta_p)^2}{4 \beta_p^2} \, 
\exp\Biggl(- \frac{4 D_\mathrm{c} \beta_p^2}{L^2} t \Biggr)
\Biggr], \\[2mm]
\lefteqn{
g_{\mathrm{AB}}(q,t) = 
\frac{N}{L^2} \int_0^{L/2} \! \int_{L/2}^{L} \phi(s,s';t) \, 
  \mathrm{d}s \, \mathrm{d}s' } \quad &
\\[4pt] & 
= N \sum_{p=1}^{\infty} \Biggl[
\frac{2 \mu}{\mu^2 + \alpha_p^2 + \mu} \, 
\frac{\sin^2 \alpha_p}{4 \alpha_p^2} \, 
\exp\Biggl(- \frac{4 D_\mathrm{c} \alpha_p^2}{L^2} t \Biggr) 
\\[4pt] & \qquad 
- \frac{2 \mu}{\mu^2 + \beta_p^2 + \mu} \, 
\frac{(1 - \cos \beta_p)^2}{4 \beta_p^2} \, 
\exp\Biggl(- \frac{4 D_\mathrm{c} \beta_p^2}{L^2} t \Biggr)
\Biggr].
    \end{aligned}
\end{equation}

 \bigskip
Using the above relations to calculate $\Delta(q,t) =
2\bigl(g_\mathrm{AA} - g_\mathrm{AB}\bigr)$ leads to Eq.~(26) of the
main manuscript.

\section{Effective coordination number in the long-chain limit}

\begin{figure}[h]
    \centering
    \includegraphics[width=0.45\textwidth]{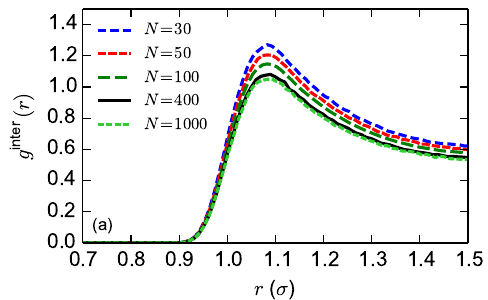}
    \includegraphics[width=0.45\textwidth]{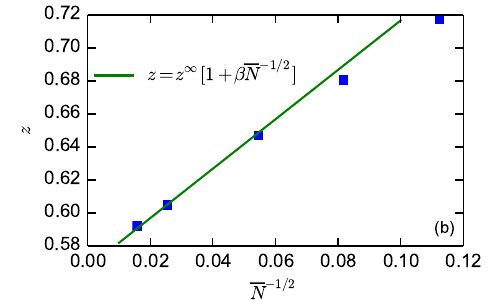}
    \caption{(a) Intermolecular radial distribution function for melts
    with different chain lengths. (b) Effective coordination number as
    a function of $\bar{N}^{-1/2}$. In this panel, solid line shows
    $z(N) = z_\infty\bigl(1 + \beta \bar{N}^{-1/2}\bigr)$ with $\beta
    = (6/\pi)^{3/2}$ and $z^\infty = 56.7$.    }
    \label{Fig:si-z} % caption for whole figure
\end{figure}

To estimate $\chi$ for a given value of the parameter $e$ ($= \varepsilon_\mathrm{AB}/\varepsilon_\mathrm{AA} - 1$), which controls the repulsion between different monomer types in the molecular dynamics simulations, we use the following relation~\cite{morse2009chain}:
\begin{equation}
    \chi(e) \approx \frac{e z_\infty}{k_\text{B}T}
    \label{si-eq:chi}
\end{equation}
where $z_\infty$ is the effective coordination number for infinitely
long chains, defined as $z_\infty = \lim_{N \to \infty} z(N) = \lim_{N
\to \infty} \int g^\mathrm{inter}(r)_N, u(r), 4\pi r^2 ,\mathrm{d}r$.
Here, $g^\mathrm{inter}(r)_N$ denotes the intermolecular radial
distribution function of a homopolymer melt, and $u(r)$ is the
non-bonded interaction potential.  To determine $z_\infty$, we
simulated homopolymer melts with varying chain lengths and calculated
the effective coordination, $z(N)$, number for each system.
\cref{Fig:si-z}a shows the intermolecular radial distribution
function, $g^\mathrm{inter}(r)$, for melts with different chain
lengths. These data are used to calculate $z(N)$. $z_\infty$ is then
determined by fitting the data to $z(N) = z_\infty\bigl(1 + \beta
\bar{N}^{-1/2}\bigr)$, with $\beta = (6/\pi)^{3/2}$, as proposed by
Morse and Chung~\cite{morse2009chain}. The quantity $\bar{N} = (\rho
R_\text{ee}^3/N)^2$ is the invariant degree of polymerization, where
$\rho$ is the monomer number density and $R_\text{ee}$ is the
end-to-end distance. 

\cref{Fig:si-z}b shows the calculated values of the effective coordination number as a function of $\bar{N}^{-1/2}$. The solid line in this panel represents the above extrapolation relation with $z_\infty = 56.7$, which shows good agreement with the data for sufficiently long chains.

\bigskip

\section{Comparison between MD and DSCFT structure factors}

\begin{figure*}[htb]
    \centering
    \includegraphics[width=0.4\textwidth]{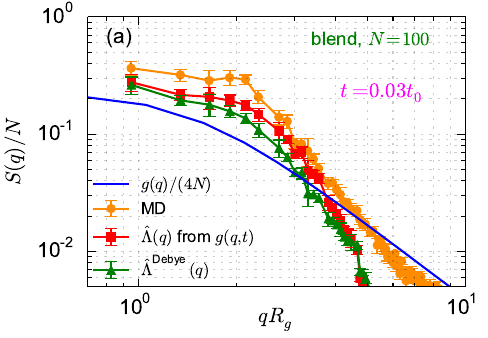}
    \includegraphics[width=0.4\textwidth]{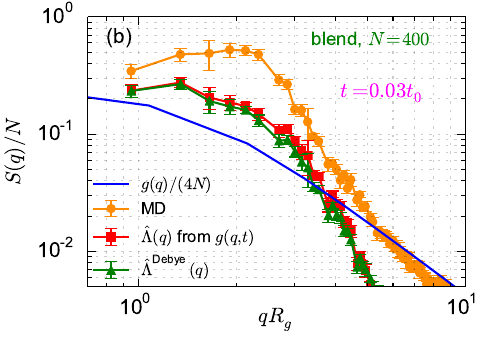}

    \includegraphics[width=0.4\textwidth]{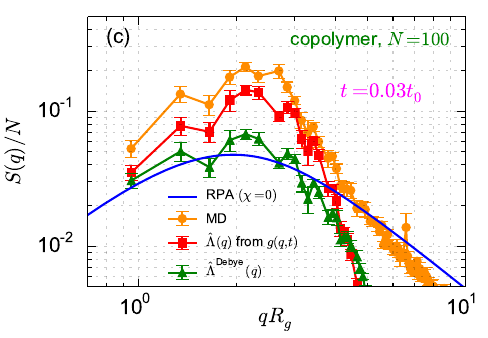}
    \includegraphics[width=0.4\textwidth]{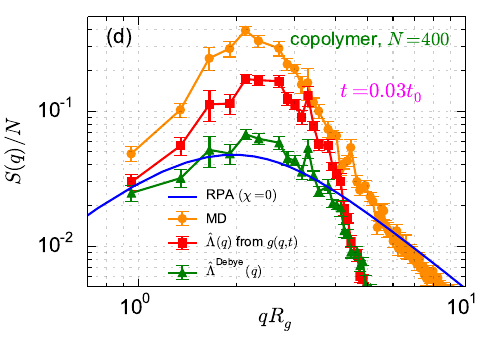}

    \caption{The normalized structure factor, $S(q)/N$, at $t = 0.03,t_0$ following a quench from $\chi N = 0$ to $\chi N = 28.3$. Panels (a) and (b) show the results for blends with chain lengths $N = 100$ and $N = 400$, respectively, while panels (c) and (d) show the results for diblock copolymers with chain lengths $N = 100$ and $N = 400$, respectively. Each panel compares the $S(q)$ obtained from MD simulations with the results from DSCFT calculations using the two mobility functions: one derived from MD simulation data for the single-chain dynamic structure factor and the other based on the Debye scheme.
   }
    \label{Fig:si-Sq}
\end{figure*}

To facilitate a clearer comparison between the MD simulation results for the dynamics of structure formation and the DSCFT results obtained using different choices of mobility functions, \cref{Fig:si-Sq} shows the normalized structure factor, $S(q)/N$, at a given time, $t = 0.03\ t_0$, following a quench from $\chi N = 0$ to $\chi N = 28.3$. The figure compares the MD results with DSCFT calculations using the two mobility functions, one derived from the single-chain dynamic structure factor and the other based on the Debye scheme, for both blends and diblock copolymers with $N = 100$ and $N = 400$.

\section{Average growth rate in RPA approximation}

The average growth rate of the structure factor is defined as
\begin{equation}
     \bar{R}_t(q) = \frac{\ln(S_t(q)/S_{t=0}(q))}{2 t}.
     \label{si-eq:Rt}
\end{equation}
(see Eq.~(31) in the main manuscript), where $S_t(q)$ denotes
the compositional structure factor at time $t$,
$S_t(\qq) = \frac{\rho_0}{V} \langle \psi(\qq,t) \psi(-\qq,t) \rangle$
with $\psi(\qq) = \frac{1}{2}
\left(\phi_\mathrm{A}(\qq)-\phi_\mathrm{B}(\qq) \right)$.
Our goal is to calculate this quantity for the process of
spinodal decomposition in RPA approximation for incompressible
systems with $\phi(\qq) = \left(\phi_\mathrm{A}(\qq)
-\phi_\mathrm{B}(\qq) \right) = V \delta_{\qq 0}$.

Our starting point is the deterministic DSCFT equation, Eq.~(4) 
in the main manuscript. After Fourier transform, it takes the form
   \begin{equation}
   \frac{\partial \phi_\alpha(\qq,t)}{\partial t}
   = -q^2 \sum_\beta \lhat_{\alpha \beta}(\qq)\:\muhat_\beta(\qq),
 \label{si-Eq:DDFT-q}
 \end{equation}
where $\lhat$ is the scaled mobility function and
$\muhat_{\beta} = V \delta F/\delta \phi_{\beta}(-\qq)$ the 
excess chemical potential in Fourier space. 
Next we perform a variable transformation from
$(\phi_\mathrm{A}, \phi_\mathrm{B})$ to $(\phi,\psi)$, and define
$\muhat_\phi(\qq) = V \delta F/\delta \phi(-\qq)$,
$\muhat_\psi(\qq) = V \delta F/\delta \psi(-\qq)$.
Expressing $\muhat_\mathrm{A}$ and $\muhat_\mathrm{B}$ 
in terms of $\muhat_\phi$ and $\muhat_\psi$ via
\begin{eqnarray*}
  \muhat_\mathrm{A,B} &=& V  \frac{\delta F}{\delta \phi_\mathrm{A,B}}
   = 
       V \left( \frac{\delta F}{\delta \phi} \:    
            \frac{\partial \phi}{\partial \phi_\mathrm{A,B}}
            +
            \frac{\delta F}{\delta \psi} \:    
            \frac{\partial \psi}{\partial \phi_\mathrm{A,B}}
      \right)
 \\  & =& \muhat_\phi \pm \frac{1}{2} \muhat_\psi,
\end{eqnarray*}
we can rewrite the dynamical equations as 
\begin{equation}
 \partial_t \left( \begin{array}{c} \phi \\ \psi \end{array} \right)
 = - q^2 \: \left( \begin{array}{cc} \lhat_{\phi \phi} & \lhat_{\phi \psi}
 \\ \lhat_{\psi \phi} & \lhat_{\psi \psi} \end{array} \right)
 \: \left( \begin{array}{c} \muhat_\phi  \\ \muhat_\psi \end{array}
 \right)
 \end{equation}
with
 \begin{eqnarray*}
 \lhat_{\phi \phi} &=&
  (\lhat_\mathrm{AA} + \lhat_\mathrm{AB} 
   + \lhat_\mathrm{BA} + \lhat_\mathrm{BB})
\\ \lhat_{\phi \psi} &=&
  \frac{1}{2} (\lhat_\mathrm{AA} - \lhat_\mathrm{AB} 
   + \lhat_\mathrm{BA} - \lhat_\mathrm{BB})
\\ \lhat_{\psi \phi} &=&
  \frac{1}{2} (\lhat_\mathrm{AA} + \lhat_\mathrm{AB} 
   - \lhat_\mathrm{BA} - \lhat_\mathrm{BB})
\\ \lhat_{\psi \psi} &=&
  \frac{1}{4} (\lhat_\mathrm{AA} - \lhat_\mathrm{AB} 
   + \lhat_\mathrm{BA} - \lhat_\mathrm{BB})
 \end{eqnarray*}
The incompressibility condition  $\partial \phi = 0$
imposes the relation 
$\muhat_\phi = - (\lhat_{\phi \psi}/\lhat) \: \muhat_\psi$
between $\muhat_\phi$ and $\muhat_\psi$.
Inserting this into the equation for $\partial \psi$,
we obtain after some algebra:
\begin{equation}
\partial_t \psi = - q^2 \:
\frac{\lhat_\mathrm{AA} \lhat_\mathrm{BB} 
      - \lhat_\mathrm{AB} \lhat_\mathrm{BA}}
      {\lhat_\mathrm{AA} + \lhat_\mathrm{AB}
       + \lhat_\mathrm{BA} + \lhat_\mathrm{BB} } 
      \: \muhat_\psi.
\end{equation}

The next task is to find an expression for the excess chemical potential
$\muhat_\psi$.  In incompressible systems, the reduced linearized 
RPA free energy is written as a function of $\psi$ 
as\cite{schmid2011}
\begin{equation}
F[\psi] = F_0 + \frac{1}{2 V} \sum_{\qq}
\Gamma_2(\qq) \: |\psi(\qq)|^2,
\end{equation}
with $\Gamma_2$ defined as in Eq.~(30) in the main text. 
Taking the derivative gives 
\begin{equation}
\muhat_\psi(\qq) = V \: \frac{\partial F}{\partial \psi(-\qq)}
 = \Gamma_2(\qq) \: \psi(\qq).
\end{equation}

Putting everything together, we finally
obtain the following dynamical equation for $S_t(\qq)$:
\begin{equation}
\partial_t S_t(\qq)
= \frac{\rho_0}{V} \langle \psi(\qq,t) \psi(\qq,-t) \rangle
=  2 \bar{R}_t(\qq) S_t(\qq)
\end{equation}
with 
\begin{equation}
\bar{R}(\qq) = - q^2 \: 
  \frac{\lhat_\mathrm{AA}(\qq) \lhat_\mathrm{BB}(\qq)
      - \lhat_\mathrm{AB}(\qq) \lhat_\mathrm{BA}(\qq)}
      {\lhat_\mathrm{AA}(\qq) + \lhat_\mathrm{AB}(\qq)
       + \lhat_\mathrm{BA}(\qq) + \lhat_\mathrm{BB}(\qq) } 
  \: \Gamma_2(\qq).
\end{equation}
In the RPA approximation, the structure factor thus
grows exponentially as expected with the $q$-dependent,
but time-independent growth rate $\bar{R}(q)$.

Applying this result to the systems considered in the main paper,
we have $\lhat_\mathrm{AA} = \lhat_\mathrm{BB} = \frac{1}{2} \lhat$
for symmetric homopolymer blends, giving
\begin{equation}
\bar{R}^\mathrm{blend}(\qq) = - q^2 \: 
  \frac{\lhat}{4} \: \Gamma_2(\qq).
\end{equation}
and $\lhat_\mathrm{AA} = \lhat_\mathrm{BB}$,
$\lhat_\mathrm{AB} = \lhat_\mathrm{BA}$,
for symmetric diblock copolymer melts, giving
\begin{equation}
\bar{R}^\mathrm{diblock}(\qq) = - q^2 \: 
  \frac{\lhat_\mathrm{AA} - \lhat_\mathrm{AB}}{2} \: \Gamma_2(\qq).
\end{equation}
These are the expressions given in the main text.

%\bibliographystyle{apsrev4-2-titles}
%\bibliography{refs}

\end{document}